\documentclass{aa}  

\usepackage{graphicx}
\usepackage{multirow}
\usepackage{txfonts}
\usepackage{natbib}
\bibpunct{(}{)}{,}{a}{}{,}  
\usepackage{orcidlink}
\usepackage{hyperref}
\hypersetup{breaklinks=true,
            colorlinks,
            filecolor=blue,
            linkcolor=blue,
            citecolor=blue,
            urlcolor=blue,
            anchorcolor=blue} 
\newcommand\feh{\ensuremath{[\mathrm{Fe}/\mathrm{H}]}}
\newcommand\afeh{\ensuremath{[\alpha/\mathrm{Fe}]}}

\newcommand\vlos{\ensuremath{V_{\mathrm{los}}}}
\newcommand\vsys{\ensuremath{V_{\mathrm{sys}}}}

\begin{document} 

   \title{Low-resolution spectroscopic characterisation of five poorly known Galactic stellar clusters}

   \authorrunning{Ceccarelli, E., et al.}   
   
   \author{E. Ceccarelli
          \inst{1,2,3}
          \orcidlink{0009-0007-3793-9766}
          \and
          M. Bellazzini
          \inst{3}
          \orcidlink{0000-0001-8200-810X}
          \and
          D. Massari
          \inst{3}
          \orcidlink{0000-0001-8892-4301}            
          \and
          A. Mucciarelli
          \inst{2,3}
          \orcidlink{0000-0001-9158-8580}                   
          \and
          M. De Leo
          \inst{2,3}
          \orcidlink{0000-0001-5844-733X}                   
          \and
          M. Libralato
          \inst{4}
          \orcidlink{0000-0001-9673-7397}       
          \and
          E. Dodd
          \inst{5}
          \orcidlink{0000-0002-2691-7728}                            
          }

   \institute{Kapteyn Astronomical Institute, University of Groningen, Landleven 12, 9747 AD Groningen, The Netherlands
             \\ \email{ceccarelli@astro.rug.nl}             
         \and 
             Dipartimento di Fisica e Astronomia "Augusto Righi", Università di 
             Bologna, Via Gobetti 93/2, 40129 Bologna, Italy
         \and
             INAF - Osservatorio di Astrofisica e Scienza dello Spazio di Bologna, Via Gobetti 93/3, 40129 Bologna, Italy 
         \and
             INAF - Osservatorio Astronomico di Padova, Vicolo
             dell’Osservatorio 5, Padova I-35122, Italy
         \and
             Institute for Computational Cosmology \& Centre for Extragalactic Astronomy, Department of Physics, Durham University, South Road, Durham, DH1 3LE, UK
             }
 
  \abstract 
  {Stellar clusters preserve crucial information on the formation and evolutionary processes that shaped the Milky Way (MW) as we see it today. However, several MW clusters still lack sufficient data to constrain their metallicity, ages, and, in some cases, even their basic kinematic properties. We present low-resolution MODS at LBT spectroscopy for five such systems (i.e. Koposov 1, Koposov 2, Mu\~noz 1, Pfleiderer 2, and RLGC2) from which we derive systemic heliocentric radial velocities ($\vsys$) with typical uncertainties of $\le10$ km s$^{-1}$, and metallicities based on the equivalent widths of the infrared Ca II triplet measured in red giant branch members. For Pfleiderer 2 and RLGC2, we provide the first spectroscopic determinations of their systemic velocities and metallicities, for which we find  $\vsys = 6 \pm 5$ km s$^{-1}$ and $-313 \pm 6$ km s$^{-1}$, and $\feh = -0.75 \pm 0.09$ dex and $-2.33 \pm 0.13$ dex, respectively. For the other three clusters, we find results that are consistent with the existing literature. Thanks to our new spectroscopic measurements, we were able to perform an orbital analysis to investigate their origin. We find that Pfleiderer 2 likely formed within the MW, RLGC2 is dynamically associated to the \textit{Gaia}–Sausage–Enceladus accretion event, and Koposov 1 was likely stripped from the Sagittarius dwarf spheroidal while Mu\~noz 1 is only tentatively associated with the latter system. Koposov 2, at a high orbital energy, does not show a clear association with any known progenitor system.}


   \maketitle

%-------------------------------------------------------------------

\section{Introduction}
Globular clusters (GCs) are among the oldest stellar systems in the Galaxy and have long been regarded as pristine fossils of its early assembly \citep{searle&zinn1978}. The discovery and characterisation of the accreted population of GCs in the Galaxy \citep[e.g.][]{forbes&bridges2010,vandenberg2013,massari19,horta2020,callingham2022,ceccarelli2024b,chen&gnedin2024,monty2024,deleo2026}, observations of massive bound clusters forming at high redshift \citep[][]{vanzella2017,claeyssens2023,adamo2024}, and the growing evidence that a large fraction of the early Milky Way (MW) may have formed within (now disrupted) GCs \citep{schiavon2017,belokurov&kravstov2023} are just a few examples that highlight their crucial role in the assembly and early chemical enrichment of galaxies. However, a significant fraction of the MW GCs system is, essentially, unexplored to date. Even though the population of massive, high-luminosity GCs in the Galactic halo is now close to being fully inventoried \citep[see e.g.][]{webb&carlberg2021}, a growing number of low-mass and low-surface-brightness stellar systems have been discovered in the last two decades \citep[e.g.][]{belokurov2010,fadely2011,minniti2011,laevens2015,kim2016,koposov2017,garro2020,gran2022}. Their physical properties often place them in a transitional regime where traditional classifications between GCs and more diffuse stellar systems (e.g. ultra-faint dwarf galaxies, open clusters) are no longer clear-cut \citep{conn2018,munoz2018,mau2020}. Moreover, many of these clusters are observationally challenging, either residing in the far reaches of the MW halo or buried behind heavy extinction towards the Galactic disc and bulge \citep{massari2025}. This means that some of the most intriguing clusters still lack even basic spectroscopic characterisation, including their systemic heliocentric radial velocity ($V_{\mathrm{sys}}$), metallicity ($\feh$), and/or $\afeh$ ratio, which limits our ability to constrain their origin, current dynamical state, and evolution. To partially address this gap, we collected low-resolution spectra for five systems observed by the \textit{Hubble} Missing Globular Cluster Survey \citep[MGCS,][]{massari2025}, namely Koposov 1, Koposov 2, Mu\~noz 1, Pfleiderer 2, and RLGC2, that are either heavily extinct and/or very sparse and distant \citep{koposov2007,ortolani2009,munoz2012,ryu2018}. By targeting the brightest accessible stars, we aim to obtain robust measurements of the clusters’ systemic velocities and, whenever possible, to derive their spectroscopic metallicity, based on the equivalent width (EW) of the infrared Ca II triplet. For Koposov 1, Koposov 2, and Mu\~noz 1, our analysis provides an independent validation of literature results \citep{munoz2012,cerny26,geha2026}, while for Pfleiderer 2 and RLGC2 we provide the first ever spectroscopic measurements. Together, these observations complement MGCS data and supply the necessary kinematic and chemical information required to place these stellar clusters in a chemo-dynamical framework, and to better constrain their role within the assembly history of the MW.

%--------------------------------------------------------------------
\section{Observations and data reduction}
\label{sec:dataset}
%--------------------------------------------------------------------
%-------------------------- Table -------------------------
\begin{table*}
\caption{Information on the observations for target clusters.}        
\label{tab:obs_clusters}      
\centering          
\begin{tabular}{lccccccc}  
\hline 
\hline      
Cluster & R.A. & Dec. & UT-Date & $t_{\mathrm{exp}}$ & Airmass & Seeing & $\mathrm{N_{\star,tot}}$ \\ 
 & (deg) & (deg) &  & (s) & & (arcsec) & \\ 
\hline 
Koposov 1 & 179.827 & 12.260 & 2025-05-03 & 2$\times$1200 & 1.0 & $1.3$ & 4  \\
Koposov 2 & 119.571 & 26.255 & 2025-01-31 & 2$\times$1200 & 1.1 & $0.9$ & 4  \\
Mu\~noz 1 & 225.450 & 66.969 & 2025-04-30 & 2$\times$1200 & 1.2 & $1.0$ & 4  \\
Pfleiderer 2 & 269.664 & -5.070 & 2025-05-23 & 3$\times$1200 & 1.4 & $1.0$ & 7  \\
RLGC2     & 281.367 & -5.192 & 2025-05-23 & 3$\times$1200 & 1.3 & $1.0$ & 4  \\
\hline                  
\end{tabular}
\tablefoot{Centre of the clusters from \citet{vasiliev&baumgardt2021}, date of the observation, exposure time ($t_{\mathrm{exp}}$), airmass, seeing, and number of stars in the mask. 
}
\end{table*}
%-------------------------------------------------------------
\begin{figure*}
\centering 
\includegraphics[width=0.95\textwidth]{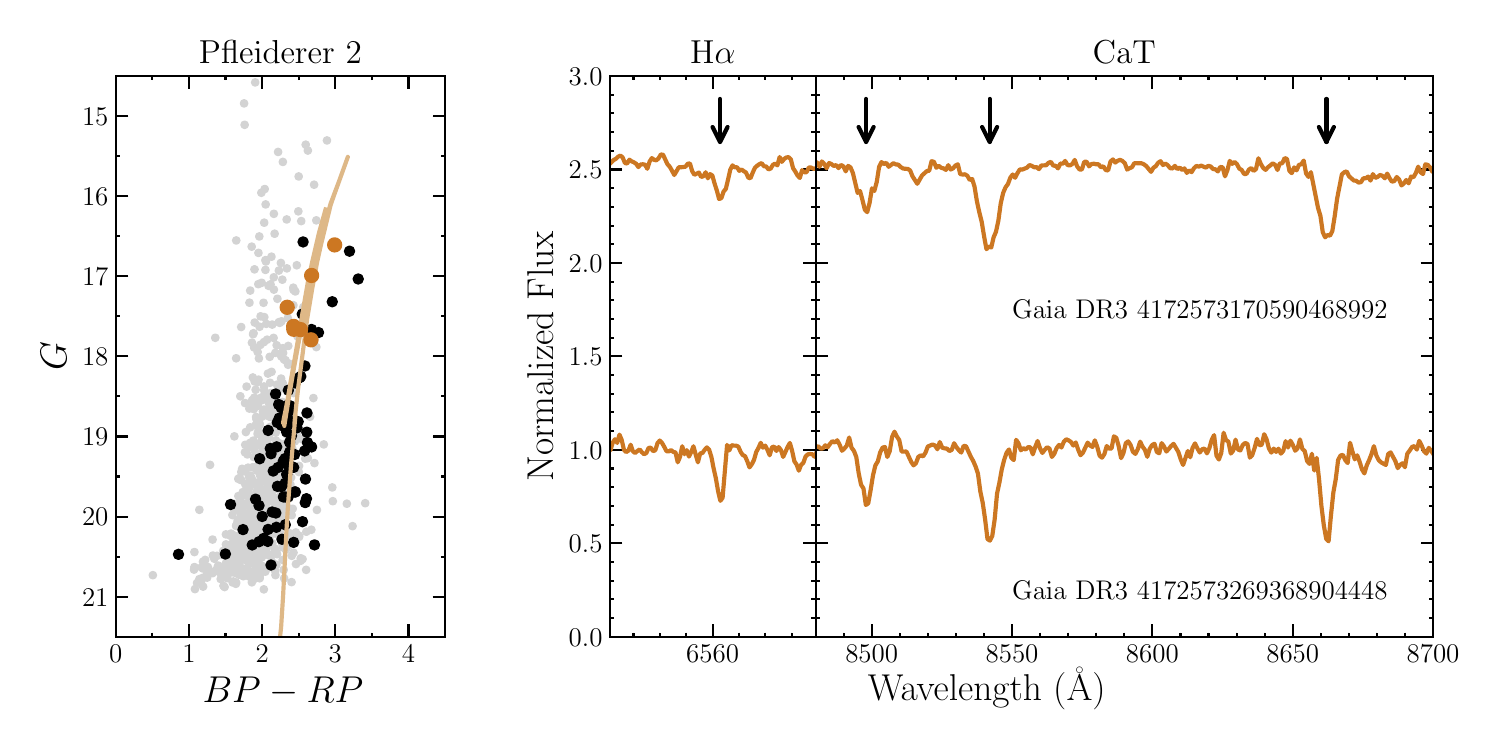}  
\caption{Left panel: \textit{Gaia} CMD for Pfleiderer 2 from \citet{vasiliev&baumgardt2021}. Typical photometric errors at these magnitudes can be found in \citet{GC21}. All stars with a probability of membership $> 80\%$ are shown in black, while our selected target stars are plotted in orange. A BaSTI isochrone \citep[][10 Gyr, $\feh = -0.7$ dex, solar-scaled $\afeh$]{hidalgo2018} is also plotted for reference, taking distance and reddening estimates from \citet{ortolani2009}. Right panel: MODS spectra for the highest and lowest S/N stars among our sample, shown around the H$\alpha$ and CaT regions (the positions of such absorption lines are indicated with arrows). The same plots for other clusters can be found in Appendix \ref{app:A} (see Figs. \ref{fig:data_2} - \ref{fig:data_3}).} 
\label{fig:data_Pfleiderer2}%
\end{figure*}
%-------------------------------------------------------------

All target clusters were observed with the low-resolution Multi-Object Double Spectrographs \citep[MODS,][]{pogge2010} at the Large Binocular Telescope (LBT, Programme IT-2024B-009, P.I.: E. Ceccarelli); see Table \ref{tab:obs_clusters} for a summary of the observations. MODS is composed of a pair of optical multi-object spectrographs (MODS1 and MODS2) mounted on the two LBT binocular arms. MODS spectra were obtained in dichroic mode using both the G400L ($3500-5900$ \r{A}) and the G670L ($5400-10000$ \r{A}) gratings on the blue and red channels, respectively. Such a wide wavelength range enables the detection of several absorption features that are key for retrieving a precise heliocentric radial velocity ($\vlos$).

For each cluster we designed a mask containing $0.6$'' slits (yielding a spectral resolution of $R\sim2300$) for up to seven stars per cluster (all the information is reported in Table \ref{tab:obs_stars}) selected according to the proper motion based membership probability ($p>80\%$) by \citet{vasiliev&baumgardt2021}. Masks were observed with MODS in dual mode with two exposures of $1200$ s each. Due to the presence of a small veil of clouds, Pfleiderer 2 and RLGC2 were observed with one extra exposure.

We reduced the spectra using the Spectroscopic Interactive Pipeline and Graphical Interface (\texttt{SIPGI}) tool \citep{gargiulo22}. In particular, we produced calibration frames for both MODS1 and MODS2 following the procedure outlined below. Firstly, we generated a bad pixel map using imaging flats and applied it to every observed frame, together with the correction for the presence of cosmic rays. We independently bias-subtracted and flat-field corrected each frame using a master flat derived from a set of multi-object spectrograph flats. We then applied the dispersion solution to each frame in order to perform the wavelength calibration and correct for optical distortions. To do so, we used a master lamp obtained from a series of arc lamp exposures. Using \texttt{SIPGI}, we were able to achieve a typical precision in the wavelength calibration of %$0.08$ \r{A}.
$\sim10$ km $\mathrm{s^{-1}}$ \citep{gargiulo22}. We then extracted the two-dimensional, wavelength-calibrated spectra and performed the sky subtraction. Finally, we extracted the one-dimensional spectrum of each source and shifted it to the air-calibrated wavelength rest-frame \citep[see e.g.][]{edlen1966,hanuschik2003}. Finally, we applied the heliocentric velocity correction along the line of sight (measured at the time of each corresponding observation with the \texttt{rvcorrect} task from \texttt{IRAF}\footnote{\texttt{IRAF} was distributed by the National Optical Astronomy Observatory (NOAO), which was managed by the Association of Universities for Research in Astronomy (AURA) under a cooperative agreement with the National Science Foundation.}) on the spectra of each star, and stacked them together to improve the signal-to-noise (S/N) ratio. 

In Fig. \ref{fig:data_Pfleiderer2} we plot the \textit{Gaia} colour-magnitude diagram (CMD) obtained from the catalogue by \citet{vasiliev&baumgardt2021} for Pfleiderer 2, together with the highest and lowest S/N stellar spectra, to show the quality of the data around some absorption features of interest, i.e. H$\alpha$ and Ca II triplet (CaT). The same plots for all clusters are in Appendix \ref{app:A}.

%-------------------------------------------------------------   
\section{Chemodynamical analysis}
\label{sec:analysis}
%-------------------------------------------------------------

In this section, we present the details of the analysis performed on the spectra. The discussion and interpretation of the results are deferred to Section \ref{sec:discussion}.

%-------------------------------------------------------------   
\subsection{Heliocentric radial velocity}
\label{sec:vlos}
%-------------------------- Table -------------------------
\begin{table*}
\caption{Photometric atmospheric parameters and results from the spectroscopic analysis.}        
\label{tab:results}      
\centering          
\begin{tabular}{lcccccc}  
\hline 
\hline      
Cluster / \textit{Gaia} DR3 & $T_{\mathrm{eff}}$ & log $g$ & $V_{\mathrm{los}}$ & EW CaT & $\mathrm{[Fe/H]_{N26,G}}$ & Member \\ 
 & (K) & (dex) & (km $\mathrm{s^{-1}}$) & (\r{A}) & (dex) &  \\ 
\hline 
3919839462184529664 & 4762 & 2.03 &     7$\pm$13 & 5.72$\pm$0.16 & $-$1.22$\pm$0.09 & y \\
3919839767126601728 & 5025 & 3.18 &    14$\pm$21 & - & - & y \\
3919839767126647680 & 6875 & 3.78 &     0$\pm$21 & - & - & y \\
3919839771422278016 & 6283 & 3.73 &    24$\pm$23 & - & - & y \\
Koposov 1 & & &   10$\pm$9 & & & \\
\hline 
874275674394169984 & 5416 & 2.65 &   113$\pm$12 & 1.65$\pm$0.13 & $-$2.91$\pm$0.12 & y  \\
874275609971796992 & 6668 & 3.60 &    90$\pm$20 & - & - & y  \\
874269730159393536 & 5517 & 3.42 &   131$\pm$21 & - & - & y  \\
874275850488194176 & 5157 & 3.42 &    26$\pm$21 & - & - & n  \\
Koposov 2 & & &  112$\pm$9 & & & \\
\hline 
1693370742839913088 &  9631 & 3.37 & $-$144$\pm$15 & - & - & y \\
1693394146117710592 &  8875 & 3.47 & $-$109$\pm$18 & - & - & y \\
1693394176181484672 &  4957 & 2.94 & $-$121$\pm$20 & 4.26$\pm$0.98  & $-$1.43$\pm$0.46 & y \\
1693393420267235840 &  5846 & 3.45 & $-$191$\pm$23 & - & - & n \\
Mu\~noz 1 & & & $-$127$\pm$10 & & &  \\
\hline 
4172573170590468992 & 3825 & 0.84 &  $-$18$\pm$13   & 8.28$\pm$0.41 & $-$0.78$\pm$0.38 & y \\
4175575730690574208 & 4810 & 1.68 &      8$\pm$12   & 7.75$\pm$0.30 & $-$0.76$\pm$0.25 & y \\
4172549561149372160 & 4207 & 1.23 &     17$\pm$12   & 7.59$\pm$0.42 & $-$0.91$\pm$0.31 & y \\
4172549836027324800 & 4625 & 1.69 &     12$\pm$12   & 7.56$\pm$0.26 & $-$0.77$\pm$0.21 & y \\
4172549840328133120 & 4621 & 1.71 &      9$\pm$13   & 8.09$\pm$0.29 & $-$0.59$\pm$0.22 & y \\
4175575863825402880 & 4219 & 1.56 &      3$\pm$13   & 7.57$\pm$0.22 & $-$0.73$\pm$0.19 & y \\
4172573269368904448 & 4450 & 1.63 &      4$\pm$14   & 7.46$\pm$0.32 & $-$0.80$\pm$0.22 & y \\
Pfleiderer 2 & & &    6$\pm$5 & & $-$0.75$\pm$0.09 & \\
\hline 
4253618683784925568 & 3479 & 0.24 & $-$315$\pm$11 & 4.40$\pm$0.13 & $-$2.30$\pm$0.33 & y \\
4253618683784927872 & 3483 & 0.46 & $-$311$\pm$13 & 4.04$\pm$0.11 & $-$2.32$\pm$0.27 & y \\
4253618683784923392 & 4089 & 0.97 & $-$320$\pm$13 & 3.91$\pm$0.12 & $-$2.31$\pm$0.24 & y \\
4253618683784929536 & 3555 & 0.60 & $-$307$\pm$12 & 3.74$\pm$0.13 & $-$2.39$\pm$0.24 & y \\
RLGC2 & & & $-$313$\pm$6 & & $-$2.33$\pm$0.13 & \\
\hline 
\end{tabular}
\tablefoot{Effective temperature, surface gravity, heliocentric radial velocity, total EW of the CaT lines, iron abundance from the \citet{navabi2026} relations (using $M_{G}$), and likely membership (according to the comparison with the BGM, see text for more details) for each star in the sample. We also report the weighted mean for $V_{\rm los}$ and [Fe/H], where the weights are given by the inverse square of the individual errors. The associated uncertainty is then computed as the inverse square root of the sum of the weights. 
}
\end{table*}
%-------------------------------------------------------------
%-------------------------------------------------------------

For each star, we measured the $\vlos$ using the \texttt{fxcor} task from \texttt{IRAF}, which implements the cross-correlation method described in \citet{tonry&davies1979}. As template spectra, we used synthetic spectra generated with the code \texttt{SYNTHE} \citep{kurucz} and adopted new \texttt{ATLAS9} model atmospheres based on the \texttt{KOALA} database \citep{mucciarelli2026}.

The atmospheric parameters used to generate the synthetic spectra were derived as follows and are listed in Table \ref{tab:obs_stars}:
\begin{itemize}
    \item Metallicity ([Fe/H]): We adopted literature metallicity estimates from isochrone fitting for Pfleiderer 2 \citep[$\feh = 0.0$ dex,][]{ortolani2009} and RLGC2 \citep[$\feh = -2.1$ dex,][]{ryu2018}. For Koposov 1, Koposov 2, and Mu\~noz 1, we adopted $\feh = -0.9$ dex, $\feh = -2.9$ dex, and $\feh = -1.4$ dex, respectively, as previously derived from spectra obtained with the DEep Imaging Multi-Object Spectrograph (DEIMOS) at Keck \citep{munoz2012,cerny26,geha2026}.  
    \item Effective temperature ($T_{\mathrm{eff}}$): We relied on colour - $T_{\mathrm{eff}}$ relations from \citet{mucciarelli21} to derive the effective temperatures using the \textit{Gaia} $(BP - RP)$ colour \citep{GaiaDR3}. 
    Since the adopted colour - $T_{\mathrm{eff}}$ relation depends on the metallicity of the star \citep{mucciarelli21}, the [Fe/H] listed above were initially assumed to estimate first-order temperature, which were subsequently refined using the spectroscopic metallicities obtained in this work (see Section \ref{sec:chemical_abu}). 
    \item Surface gravity (log $g$): We estimated the log $g$ starting from the Stefan-Boltzmann relation; we adopted the $T_{\mathrm{eff}}$ obtained photometrically and assumed a stellar mass of $M_{\star} = 0.8\; M_{\odot}$, which is representative for evolved stars in old GCs, as indicated by theoretical stellar evolution models \citep{hidalgo2018,pietrinferni2021}. 
\end{itemize}

For the stars with a S/N $> 10$ around both the H$\alpha$ line and the CaT lines, the $\vlos$ was derived through a single cross-correlation with a synthetic spectrum with proper atmospheric parameters, performed over the full wavelength range and masking the portions of the spectrum affected by telluric absorptions. However, this approach proved problematic for the two brightest stars in Mu\~noz 1. As horizontal branch stars (see Fig. \ref{fig:data_3}), their spectra are heavily contaminated by hydrogen Paschen lines within the CaT region, which compromise the velocity fit. Therefore, for these two stars and for the remaining targets, we only used the H$\alpha$ line, as the combination of the low S/N ratio and/or the intensity of the CaT lines was insufficient to provide reliable measurements (see Table \ref{tab:obs_stars}). The resulting values are reported in Table \ref{tab:results}.  The uncertainty on the measured $V_{\mathrm{los}}$ was computed by accounting for two independent contributions. The first term accounts for the precisions achieved with the wavelength calibration (see Section \ref{sec:dataset}). The second term accounts for the quality of the spectrum (i.e. S/N) and is derived from Monte Carlo simulations. For each star, we generated 500 synthetic spectra, adding Poisson noise to match the average S/N of the observed MODS spectrum, and resampled them to the instrumental pixel scale. Then, we computed the cross-correlation functions between these noisy spectra and the corresponding synthetic spectrum using the same procedure adopted for the data (including telluric masking when applicable or restricting the analysis to the H$\alpha$ region for lower S/N spectra). The $1\sigma$ dispersion of the resulting $V_{\mathrm{los}}$ distribution was taken as this contribution to the uncertainty. These terms were combined in quadrature to obtain the final uncertainty. Final values are reported in Table \ref{tab:results}, together with the average systemic heliocentric radial velocity calculated from likely members (see following discussion). We note that we have six stars in common with \citet{geha2026}, two in Koposov 1, three in Koposov 2, and one in Munoz 1, and the $\vlos$ we measure for these stars are always consistent within the uncertainties. 

\begin{figure*}
\centering
\begin{minipage}{0.45\textwidth}
\centering        
\includegraphics[width=0.8\textwidth]{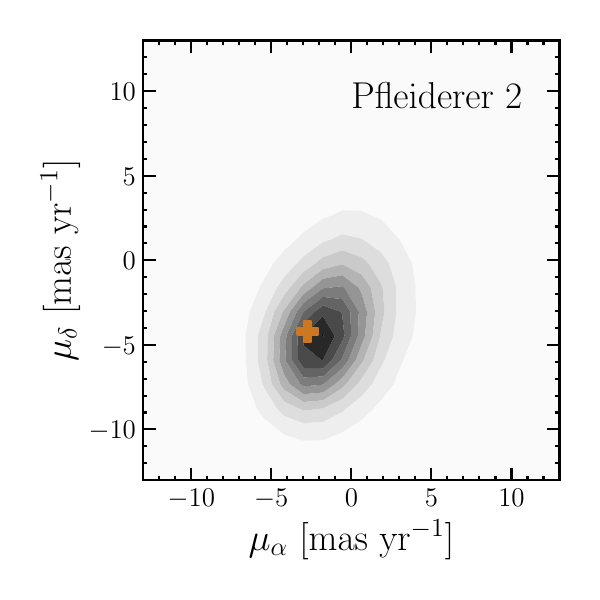}  
\includegraphics[width=1.0\textwidth]{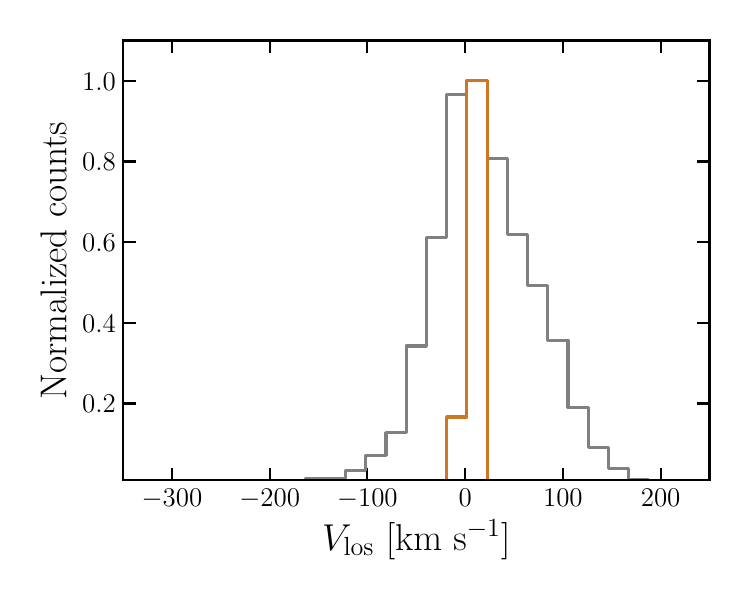}           

\end{minipage}
\begin{minipage}{0.45\textwidth}
\centering
\includegraphics[width=1.0\textwidth]{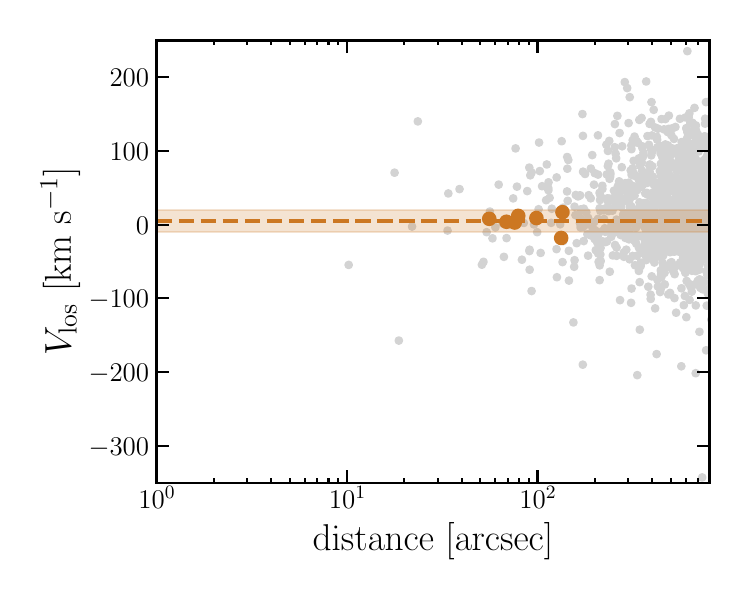} 
\includegraphics[width=1.0\textwidth]{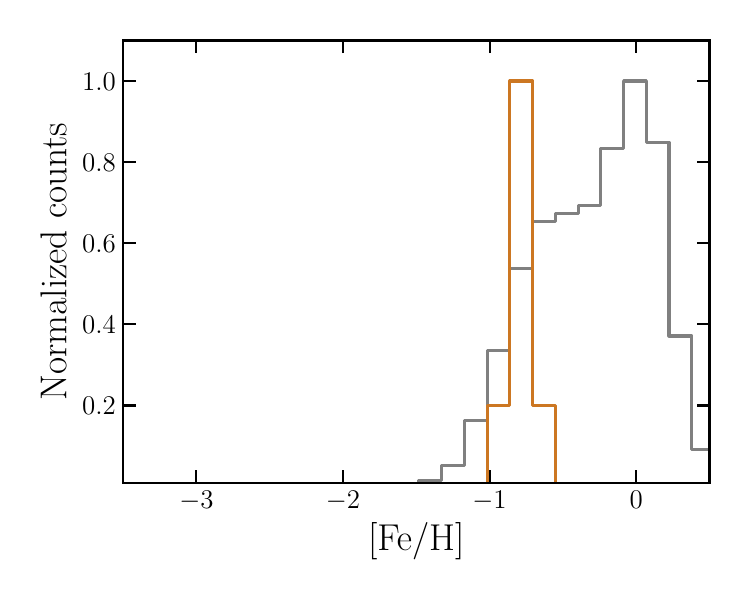}     
\end{minipage}  

\caption{Top row: Vector-point diagram (left) for a synthetic Galactic field population (located within $1^\circ$ from Pfleiderer 2 and selected in the same CMD region populated by the RGB of the cluster) computed with a BGM \citep[][]{robin2003}. The proper motion of Pfleiderer 2 from \citet{vasiliev&baumgardt2021} is indicated with a $\lq$plus' marker. On the right panel we display the $\vlos$ distribution of cluster (orange) and field stars from the BGM (grey) as a function of the distance from the cluster centre. Here, stars identified as outliers in $\vlos$ are marked with crosses. Bottom row: Heliocentric radial velocity (left) and metallicity (right) distributions for Pfleiderer 2 (orange) and for the BGM synthetic field population (grey).}
\label{fig:besa_pwm2}%
\end{figure*}

To interpret the results, we compared the kinematics (and metallicity, see Section \ref{sec:chemical_abu}) of the observed stars with synthetic Galactic field populations, computed with the Besan\c{c}on Galaxy Model\footnote{\url{https://model.obs-besancon.fr}} \citep[BGM,][]{robin2003}. The comparison samples include stars located within $1^\circ$ of the centre of each cluster and selected in regions of the CMD populated by our stellar targets. We show the results for Pfleiderer 2 in Fig. \ref{fig:besa_pwm2} and for all other targets in Figs. \ref{fig:app_besa1} - \ref{fig:app_besa2}. For three of the clusters (Koposov 1, Pfleiderer 2, and RLGC2), we find that all targeted stars have heliocentric radial velocities that are consistent among each other within the uncertainties. For Pfleiderer 2 and Koposov 1, the $\vsys$ we measure coincide with the peak of the velocity distribution predicted for the field population by the BGM, and their mean proper motions \citep[see][]{vasiliev&baumgardt2021} are also partially consistent with those of the synthetic field population. Nevertheless, there is an extremely low probability of randomly drawing  a number of stars equal to those that define our sample from the BGM distributions, assuming Gaussian uncertainties comparable to the typical measurement errors and simultaneously matching the three-dimensional kinematic properties of the two clusters within the observed ranges ($p\sim3\times10^{-28}$ and $p\sim1\times10^{-12}$ for Pfleiderer 2 and Koposov 1, respectively). In contrast, RLGC2 lies well outside both the $\vlos$ and proper motion distributions expected for the field population, which means that virtually no field star is plausibly expected to contaminate our sample. These comparisons strongly indicate that targeted stars are bona fide members of their respective clusters. For both Koposov 2 and Mu\~noz 1, we find that one out of the four observed stars has a $\vlos$ that is inconsistent with the average of the other three targets (at $3.7\sigma$ and $2.5\sigma$, respectively, see hatched histograms in Figs. \ref{fig:app_besa1} - \ref{fig:app_besa2}). We note that only $\sim 1\%$ of the respective surrounding Galactic field populations have proper motions compatible with those of the clusters. In the case of Koposov 2, the $\vsys$ we measure for the cluster is also offset from the peak of the field-star velocity distribution predicted by the BGM by $\sim100$ km $\mathrm{s^{-1}}$. Interestingly, the star with a different $\vlos$ instead falls close to that peak, which makes it a possible candidate field contaminant. For Mu\~noz 1, the situation is somewhat different: the cluster systemic velocity lies within the bulk of the field $\vlos$ distribution. Nevertheless, the probability of drawing a contaminant with a velocity comparable to that of the outlier star is only $p\sim1\times10^{-3}$. Despite this, Mu\~noz 1 is an ultra-faint stellar cluster for which a very small intrinsic velocity dispersion is expected \citep{cerny26}; therefore, we conservatively exclude this star from the cluster sample. These stars have been marked with a coloured cross in Figs. \ref{fig:data_2}, \ref{fig:data_3}, \ref{fig:app_besa1}, and \ref{fig:app_besa2}. The derived $\vsys$ of Koposov 1, Koposov 2, and Mu\~noz 1 are measured as the average of the $\vlos$ of likely members. We report values of $\vsys = 10 \pm 9$ km $\mathrm{s^{-1}}$, $\vsys = 112 \pm 9$ km $\mathrm{s^{-1}}$, and $\vsys = -127 \pm 10$ km $\mathrm{s^{-1}}$ , respectively, which are fully consistent with those reported in the literature \citep{munoz2012,geha2026}.

%---------------------------------------------------------------
\subsection{Orbital parameters}
\label{sec:op}
%-------------------------------------------------------------
%-------------------------------------------------------------
%---------------------------------------------------

Thanks to the spectroscopic determination of the $\vsys$, we are now able to accurately investigate the motion of these clusters within the Galaxy\footnote{As such measurements were previously unavailable for all of our targets except Mu\~noz 1, previous works inferred their orbital properties by adopting a flat prior on the $\vsys$ \citep[e.g.][]{callingham2022}.}. In particular, we complemented the $V_{\rm sys}$ measurements with the full five-dimensional phase-space information, taking positions and proper motions from \textit{Gaia} data \citep{vasiliev&baumgardt2021} and distances from isochrone fitting \citep{ortolani2009,munoz2012,paust2014,ryu2018}. We note that, for some of these clusters, the distance estimates reported in the literature differ by up to $\sim 15$ kpc \citep[see e.g.][]{koposov2007,paust2014,munoz2018,cerny26,geha2026}. We therefore discuss in Appendix \ref{app:dist} how the adopted distance influences the inferred dynamical associations and the derived metallicities. For Koposov 2, Pfleiderer 2, and RLGC2, we adopted proper motions obtained from the combination of \textit{Gaia} and \textit{HST} data as provided by the MGCS survey, which allowed measurements with improved precision compared to \textit{Gaia} alone \citep{libralato2026}. All the values are listed in Table \ref{tab:op_input}.
   
\begin{figure*}
\centering
\begin{minipage}{0.45\textwidth}
\centering        
\includegraphics[width=1.0\textwidth]{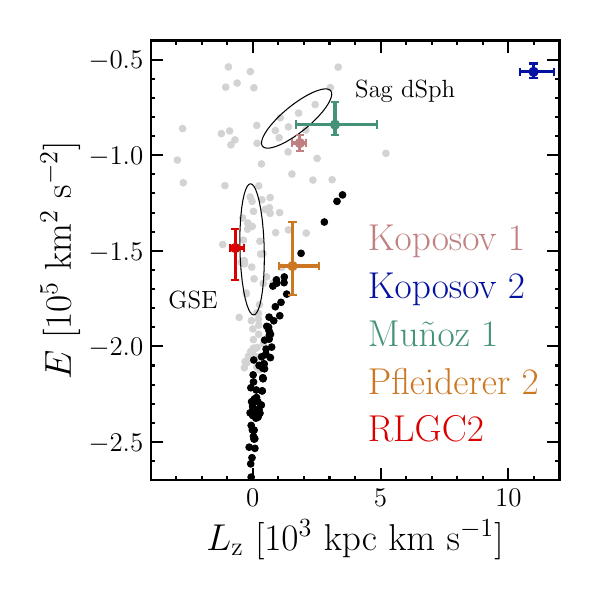}           
\end{minipage}
\begin{minipage}{0.45\textwidth}
\centering
\includegraphics[width=1.0\textwidth]{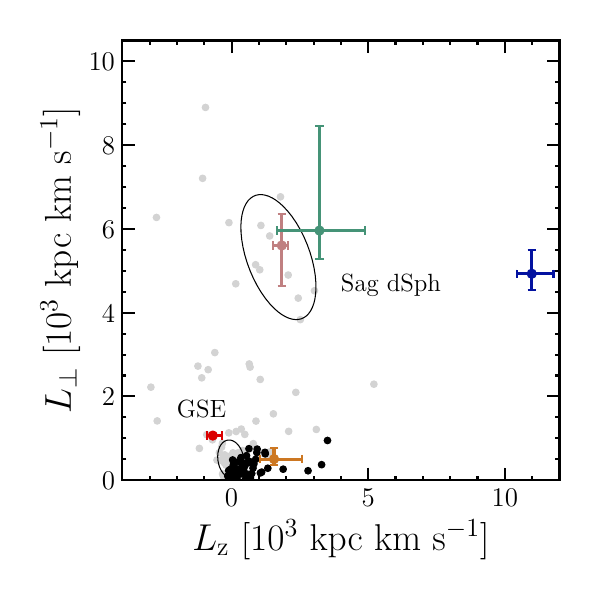}     
\end{minipage}  
\caption{Distribution in the integral-of-motions ($E, L_{\mathrm{z}}, L_{\perp}$) space of the stellar clusters studied in this paper, compared to all MW GCs (grey points) from \citet[][eDR3 edition]{massari19}. As reference, we plot the $1\sigma$ ellipses of the distributions of GCs dynamically associated with GSE and Sag dSph, while in-situ GCs are plotted in black. Error bars represent statistical uncertainties from the orbit integration.}
\label{fig:op}%
\end{figure*}

To compute the cluster orbits, we adopted a Galactocentric reference frame assuming the solar position and motion to be those defined in \citet{drimmelandpoggio18}, \citet{gravitycollaboration18}, and \citet{bennetandbovy19}. We integrated orbits within the \citet{mcmillan17} MW potential using the \texttt{AGAMA} package \citep{vasiliev19}. For each object, we performed $100$ Monte Carlo orbit realisations, assuming Gaussian uncertainties in distance, proper motion, and systemic heliocentric radial velocity. Final orbital parameters were taken as the median of the resulting distributions, with associated $\pm1\sigma$ uncertainties. Figure \ref{fig:op} illustrates the location of the clusters in the $E$ - $L_{\rm z}$ - $L_{\perp}$ space, compared with the distribution of the full MW GC population \citep[][eDR3 edition]{massari19}\footnote{See \citet{massari25} for details.}. To guide the eye, we colour in-situ GCs in black, and we draw two ellipses that highlight the $1\sigma$ distributions of GCs formed in the \textit{Gaia}-Sausage-Enceladus dwarf galaxy \citep[GSE,][]{belokurov2018,helmi2018} or in the Sagittarius dwarf spheroidal \citep[Sag dSph,][]{ibata94}, according to \citet[][eDR3 edition]{massari19}. A detailed discussion of the associations of these clusters is presented in Section \ref{sec:discussion}.

To assess the impact of the rotating bar of the MW on the recovered dynamical parameters (and thus on the associations), we redid the integration with the code \textsc{OrbIT} \citep{deleo2026_bo,deleo2026}. We found that the introduction of the bar in the underlying potential used for the integration produces minimal changes ($<10\%$) in the dynamical parameters of most of the target clusters, and does not alter the associations. Pfleiderer 2 is the exception, being the most impacted by the action of the bar, with changes in the orbital parameters of up to about $\sim30\%$ (see Section \ref{sec:discussion}).

%-------------------------------------------------------------
\subsection{Metallicity from CaT equivalent widths}
\label{sec:chemical_abu}
%-------------------------------------------------------------
%-----------------
\begin{table*}
\caption{Five-dimensional phase-space information (position, proper motion, and distance) from the literature for the target clusters.}
\label{tab:op_input}
\centering
\begin{tabular}{cccccc}
\hline\hline 
Cluster & R.A. & Dec. & $\mu_{\alpha}\rm cos(\delta)$ & $\mu_{\delta}$ & Distance \\
 & (deg) & (deg) & ($\rm mas \; yr^{-1}$) & ($\rm mas \; yr^{-1}$) & (kpc) \\
\hline
Koposov 1     & $179.827$ [1] & $12.260$ [1] & $-1.513\pm0.135$ [1] & $-0.814\pm0.105$ [1] & $33.3\pm1.5$ [2]  \\
Koposov 2     & $119.571$ [1] & $26.255$ [1] & $-0.745\pm0.056$ [3] &  $0.140\pm0.042$ [3] & $34.9\pm1.6$ [2] \\
Mu\~noz 1     & $225.450$ [1] & $66.969$ [1] & $-0.100\pm0.203$ [1] & $-0.020\pm0.207$ [1] & $45.0\pm5.0$ [4]  \\
Pfleiderer 2  & $269.664$ [1] & $-5.070$ [1] & $-2.722\pm0.018$ [3] & $-4.190\pm0.018$ [3] & $16.0\pm2.0$ [5]  \\
RLGC2         & $281.367$ [1] & $-5.192$ [1] & $-2.315\pm0.033$ [3] & $-1.864\pm0.027$ [3] & $15.8\pm2.4$ [6]  \\
\hline 
\end{tabular}
\tablefoot{The $\vlos$ is reported in Table \ref{tab:results}. Numbered references are: [1] \citet{vasiliev&baumgardt2021}, [2] \citet{paust2014}, [3] \citet{libralato2026}, [4] \citet{munoz2018}, [5] \citet{cerny26}, and [6] \citet{ryu2018}.}
\end{table*}
%-----------------
%-----------------
\begin{table*}
\caption{Orbital parameters for the observed clusters obtained in a \citet{mcmillan17} potential for the MW.}
\label{tab:op}
\renewcommand{\arraystretch}{1.5}
\centering
\begin{tabular}{lccccccc}
\hline\hline 
Cluster & Energy & $L_{\rm z}$ & $L_{\rm perp}$ & $r_{\rm peri}$ & $r_{\rm apo}$ & $z_{\rm max}$ & circ \\
 & ($\times10^{5} \; \rm km^{2} \; s^{-2}$)  & \multicolumn{2}{c}{($\times10^{3} \; \rm kpc \; km \; s^{-1}$)} & (kpc) & (kpc) & (kpc) &  \\
\hline
Koposov 1    & $-0.94^{+0.04}_{-0.04}$ &  $1.83^{+0.23}_{-0.31}$ & $5.60^{+0.76}_{-0.96}$ & $21.15^{+5.11}_{-5.21}$ & $37.63^{+1.67}_{-1.66}$ & $34.54^{+1.73}_{-2.14}$ & $0.28^{+0.01}_{-0.02}$  \\
Koposov 2    & $-0.56^{+0.04}_{-0.04}$ & $10.99^{+0.79}_{-0.55}$ & $4.93^{+0.57}_{-0.40}$ & $39.00^{+1.97}_{-1.93}$ & $107.99^{+18.26}_{-10.90}$ & $37.39^{+8.20}_{-4.02}$ & $0.73^{+0.02}_{-0.03}$ \\
Mu\~noz 1    & $-0.84^{+0.12}_{-0.05}$ &  $3.22^{+1.65}_{-1.55}$ & $5.96^{+2.49}_{-0.68}$ & $26.88^{+15.21}_{-9.78}$ & $49.39^{+11.98}_{-5,43}$ & $40.81^{+9.14}_{-5.86}$ & $0.40^{+0.07}_{-0.16}$ \\
Pfleiderer 2 & $-1.58^{+0.23}_{-0.15}$ &  $1.55^{+1.03}_{-0.51}$ & $0.50^{+0.27}_{-0.15}$ &  $5.81^{+4.37}_{-2.21}$ & $9.75^{+4.64}_{-1.86}$ & $2.88^{+1.10}_{-0.41}$ & $0.84^{+0.07}_{-0.10}$ \\     
RLGC2        & $-1.49^{+0.16}_{-0.10}$ & $-0.69^{+0.33}_{-0.22}$ & $1.06^{+0.03}_{-0.02}$ &  $2.28^{+0.55}_{-0.57}$ & $14.89^{+3.52}_{-4.69}$ & $8.21^{+7.29}_{-5.22}$ & $-0.31^{+0.18}_{-0.25}$ \\            
\hline 
\end{tabular}
\end{table*}
%-----------------

To measure the EW for each CaT absorption feature (8498.0, 8542.1, and 8662.1 \r{A}), we fitted each line in the normalised spectrum with a Voigt function, thereby accounting for the non-Gaussian profile of their wings. The line centres were fixed to their nominal wavelengths, while all other parameters in the Voigt model were left free during the fit. The EWs were computed by numerically integrating the best-fitting model over the line region. Uncertainties were estimated via Monte Carlo simulations by repeating the measurements over 500 synthetic spectra obtained by adding Poisson noise (according to the S/N of each star). The final EW uncertainty was defined as the $1\sigma$ confidence interval of the resulting EW distribution.

To convert EWs into metallicities, we used the recently published calibrations by \citet{navabi2026}, which use the absolute \textit{Gaia} $G$ magnitude as a luminosity proxy. This choice is driven by the fact that, for the clusters analysed in this work, it is not possible to reliably estimate the apparent magnitude of the horizontal branch due to the lack of multiple stars in this evolutionary sequence. Robust average metallicity estimates based on multiple red giant branch (RGB) stars are only possible for Pfleiderer 2 and RLGC2. For Koposov 1, Koposov 2, and Mu\~noz 1, our sample included only a single RGB star with sufficient S/N to derive a reliable metallicity. We note that the two brightest stars in Mu\~noz 1 are significantly hot ($T_{\rm eff} > 8800$ K) and their CaT spectral region is contaminated by hydrogen Paschen lines (see Fig. \ref{fig:data_3}), which prevents a reliable metallicity determination. Also, the star used for Mu\~noz 1 is fainter than $M_{V} = 2$, and therefore the adopted relations were applied beyond their nominal calibration range. The uncertainties were derived by propagating the errors associated with the EW measurements, the observed magnitudes and distances, as well as the intrinsic uncertainty of the calibration relation \citep[][]{navabi2026}. All the values are listed in Table \ref{tab:results}. In Appendix \ref{app:met}, we compare the values derived from the \citet{navabi2026} calibration with those obtained using alternative relations \citep{carrera2007,starkenburg2010,carrera2013}. Such an exercise is aimed at assessing the robustness of the metallicity estimates (see Table \ref{tab:app_met}). 

All three stars in Koposov 1, Koposov 2, and Mu\~noz 1 for which we are able to measure the metallicity are included in the catalogue published by \citet{geha2026}. For these objects, we find excellent agreement with their reported values, which are $\feh = -1.02 \pm 0.15$ dex, $\feh = -2.89 \pm 0.18$ dex, and $\feh = -1.26 \pm 0.27$ dex, respectively. Furthermore, if we consider the cluster metallicities derived by \citet{cerny26}, we still obtain values that are consistent with our estimates within the quoted uncertainties for Koposov 2 and Mu\~noz 1, while we find Koposov 1 to be more metal poor by $\sim 0.4$ dex, and not consistent within $2.7\sigma$. When compared with the metallicity distribution of field stars in the surrounding regions of the BGM, we note that only a handful of field stars around Koposov 2 reach such low metallicities ($< 1\%$ with $\feh < -2.5$ dex, see Fig. \ref{fig:app_besa1}). Given that Koposov 2 is observed close on the sky to the Sag dSph, it is worth noting that stars with $\feh < -2.0$ dex are known to exist in the Sag dSph, although they constitute only a very small fraction of its stellar population \citep[e.g.][]{mucciarelli2017,hayes2020,minelli23}. For Koposov 1, Mu\~noz 1, and Pfleiderer 2 the probability of finding field stars with comparable metallicities is not negligible, although the average values of the clusters remain offset from the peak of the field stars metallicity distribution. This should nevertheless be considered together with the fact that the likelihood of a field star simultaneously matching the observed proper motion and $\vsys$ is already low (see Section \ref{sec:vlos}). Finally, RLGC2 is $\sim1$ dex below the most metal-poor tail of the metallicity distribution of the surrounding stellar population. Therefore, we deem these stars as probable members of each respective cluster.

%---------------------------------------------------------------
\section{On the origin of target clusters}
\label{sec:discussion}
%---------------------------------------------------------------
%-------------------------------------------------------------
%-------------------------------------------------------------
%---------------------------------------------------------------
\subsection{Koposov 1 and Koposov 2}
%---------------------------------------------------------------
Koposov 1 and Koposov 2 are remote halo clusters located near the leading arm of the Sagittarius stream \citep{belokurov2006,koposov2007}, and have therefore been proposed as potential members of the Sagittarius dwarf spheroidal galaxy \citep{paust2014,bellazzini2020}. The association of these clusters with Sagittarius, however, has been long debated based on their dynamical properties \citep[e.g.][]{massari19,callingham2022}. In particular, \citet[][eDR3 edition]{massari19} linked Koposov 1 to the Cetus stream \citep{newberg2009ApJ}, while \citet{callingham2022} assigned it to Sagittarius with a probability of $70\%$. By incorporating our improved spectroscopic determination of the systemic $\vsys$, and following the classification scheme of \citet[][eDR3 edition]{massari19}, we find Koposov 1 to be dynamically consistent with the Sagittarius system, as Cetus stars have typically larger $L_{\mathrm{z}}$ and $L_{\perp}$, as well as lower binding energy. In contrast, Koposov 2 does not appear to be associated with any of the major accreted substructures and is therefore classified as high-energy and/or unassociated, in agreement with previous works in the literature \citep[e.g.][]{callingham2022,libralato2026}.

Early photometric analyses based on isochrone fitting \citep{koposov2007,paust2014} suggested that both systems are relatively young ($\sim5-8$ Gyr). However, no consensus was reached regarding either their distances (see Appendix \ref{app:dist}) or their metallicity, as they were alternatively reported as either metal poor ($\feh \sim -2.0$ dex) or  metal rich ($\feh \sim -0.6$ dex). This is probably due to the very low number of RGB stars in their CMDs, which prevents a good metallicity constraint for the fit, as well as the presence of potential contamination from the Sag dSph field.

Our spectroscopic analysis, in agreement with the recent results of \citet{geha2026}, places Koposov 1 at an intermediate metallicity ($\feh \sim -1.2$ dex). This metallicity is consistent with the peak of the metallicity distribution of the leading arm of the Sag dSph \citep{hayes2020}, and $\sim 0.8$ dex more metal rich than the average $\feh$ of Cetus stars \citep[see e.g.][]{sitnova2024}. Therefore, that Koposov 1 is a member of the Sag dSph system is supported by various forms of independent evidence.

We confirm the extremely metal-poor nature of the RGB star in Koposov 2 \citep[$\feh \sim -2.9$ dex, see][]{geha2026}, which would make it one of the most metal-poor stellar clusters known in the Galaxy. Surviving GCs in the Galactic halo are observed down to $\feh \sim -2.4$ dex, while even lower metallicities ($\feh < -3$ dex) have been inferred for disrupted systems \citep[see e.g.][]{martin2022}. The recent analysis by \citet{cerny26} further supports the identification of the observed RGB star as a bona fide cluster member, revising the cluster’s properties by fitting its CMD with a 13.5 Gyr old and $\feh = -2.2$ dex isochrone, and inferring a heliocentric distance of 24 kpc. In this context, Koposov 2 could in principle be interpreted as a candidate MW satellite galaxy, as systems of this kind may host a very metal-poor RGB star and possibly show hints of an intrinsic velocity dispersion, as tentatively suggested by \citet{cerny26}. Therefore, Koposov 2 matches some of the properties of the proposed class of globular-cluster-like dwarfs \citep{taylor2025}, systems thought to have formed in low-mass dark matter halos through a single, self-quenching star-formation episode at very high redshift. However, the very limited number of stars together with the current large uncertainties on the membership, distance, metallicity, and age of this system leave its nature an open question.

%---------------------------------------------------------------
\subsection{Mu\~noz 1}
%---------------------------------------------------------------
Mu\~noz 1 is an ultra-faint stellar cluster originally identified as an overdensity within the tidal radius of the Ursa Minor dwarf spheroidal galaxy, yet its significantly different heliocentric distance and $\vsys$ exclude any possible association with Ursa Minor \citep{munoz2012,munoz2018}. Both \citet[][eDR3 edition]{massari19} and \citet{callingham2022} associate Mu\~noz 1 with the Sagittarius dwarf spheroidal galaxy on the basis of its orbital properties. We find that its integrals of motion lie close to the characteristic locus of stars and clusters in the Sag dSph, which suggests a possible association, but the cluster is far from the great circle traced by the Sagittarius Stream in the sky \citep{bellazzini2020,vasiliev2021}. A first isochrone fit, based on the limited number of stars populating the RGB, indicates that Mu\~noz 1 hosts an old stellar population ($\sim12.5$ Gyr) with $\feh=-1.5$ dex \citep{munoz2012}. This metallicity estimate has subsequently been confirmed by spectroscopic observations \citep{cerny26,geha2026} and is consistent with the value we measure. 
%---------------------------------------------------------------
\subsection{Pfleiderer 2}
%---------------------------------------------------------------
%-------------------------------------------------------------
\begin{figure*}
\centering
\begin{minipage}{0.45\textwidth}
\centering        
\includegraphics[width=1\textwidth]{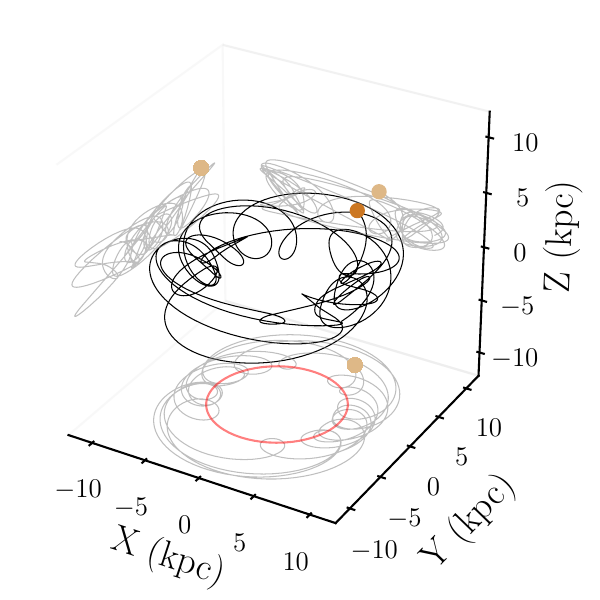}           
\end{minipage}
\begin{minipage}{0.45\textwidth}
\centering
\includegraphics[width=1\textwidth]{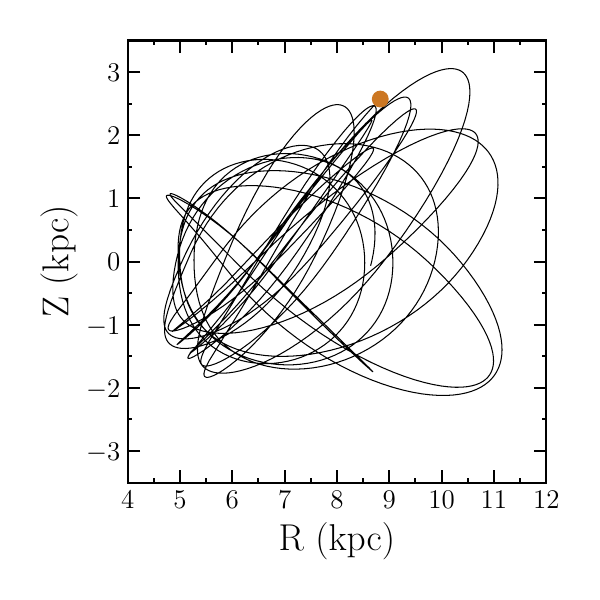}     
\end{minipage}  
\caption{Orbit of Pfleiderer 2 integrated backwards in time for 2.5 Gyr using \textsc{OrbIT} \citep{deleo2026_bo} and accounting for the presence of the bar of the MW. The current position of the cluster is shown with an orange point. The red circle shown in the X-Y projection of the orbit indicates the co-rotational radius, which is equal to 5.79 kpc for the assumed potenital in \textsc{OrbIT}.}
\label{fig:orbit_pwm2}%
\end{figure*}
%-------------------------------------------------------------
Pfleiderer 2 is a relatively old ($10 \pm 2$ Gyr) cluster that, based on isochrone fits to its tilted RGB and red horizontal branch, was initially proposed to be among the most metal-rich systems in the Galaxy \citep[$\feh \sim 0.0$ dex,][]{ortolani2009}. Its position $>2$ kpc above the Galactic plane makes it a rare example of a metal-rich cluster in such a location, comparable to only a few in-situ systems with similar properties \citep[e.g. Palomar 8 and Palomar 11,][]{harris2010}.

As shown in Table \ref{tab:results}, all seven target stars of Pfleiderer 2 have compatible $\vlos$ and metallicity. We find $\vsys = 6\pm5$ km $\mathrm{s^{-1}}$, and we revise its metallicity downwards, deriving a spectroscopic value of $\feh = -0.75\pm0.09$ dex. Given the large uncertainties on the $\vlos$, we are not able to resolve the intrinsic velocity dispersion of this cluster. \citet[][eDR3 edition]{massari19} propose this cluster to have formed in-situ, while \citet{callingham2022} assign almost equal probabilities to be associated with either GSE and the progenitor of the MW. We find that its orbital parameters are consistent with those of in-situ clusters, as it moves on a mildly heated disc-like orbit (see Table \ref{tab:op}), and is currently observed near its apocentre and maximum height over the midplane. Moreover, its relatively high metallicity argues against an accreted origin.

Our tests with \textsc{OrbIT} have shown that this is the system most impacted by the introduction of the rotating bar in the MW potential, with changes in the orbital parameters of up to $\sim30\%$. Such a big impact from the introduction of the bar prompted us to investigate further the orbital characteristics of Pfleiderer 2, following the method presented in \citet{deleo2026_bo} to check whether the cluster might be trapped on a resonant orbit. Briefly, the method is based on a series of checks on the orbital parameters (mainly pericentre and eccentricity), characteristic energy and angular momentum \citep{moreno2015, moreno2021}, and orbital frequencies \citep{bt2008, portail2015, queiroz2021} to assess if a tracer lies inside resonant loci. Our analysis suggests that Pfleiderer 2 may be on a resonant orbit when accounting for the observational uncertainties, particularly in its velocity vector. However, as the cluster lies near the limits of the adopted tolerance criteria, this trapping is likely to be weak. This is supported by Fig. \ref{fig:orbit_pwm2}, where we show the projection of the non-regular orbit of Pfleiderer 2 on the three Galactocentrc cartesian planes (X-Y, X-Z, Y-Z, left panel) and in the R-Z space (right panel). The red circle in the X-Y projection is the co-rotational radius of the bar ($5.79$ kpc for the potential adopted in \textsc{OrbIT}) and the loops of the orbit of the cluster around it show that it might be trapped near the co-rotation resonance.

%---------------------------------------------------------------
\subsection{RLGC2}
%---------------------------------------------------------------
RLGC2 is a metal-poor cluster located in the thick disc of the MW \citep{ryu2018}. Since the available photometry was not deep enough to observe the main sequence and/or sub-giant branch regions, its metallicity has been previously derived via isochrone fitting by fixing its age to the average value of that of metal-poor clusters in the MW \citep{vandenberg2013}. 

As for Pfeiderer 2, all stars in RLGC2 in our sample have compatible spectroscopic $\vlos$ and $\feh$, yielding average values of $\vsys =-313\pm6$ km $\mathrm{s^{-1}}$ and $\feh = -2.33\pm 0.13$ dex, which is $\sim -0.2$ dex more metal poor than previous photometric estimates \citep{ryu2018}. We find that RLGC2 becomes consistent with the hypothesis of being previously associated with the GSE dwarf galaxy, since it moves on a retrograde and highly eccentric orbit ($ecc \sim 0.7$) and is therefore only currently crossing the MW thick disc. This result agrees with previous classifications by \citet[][eDR3 edition]{massari19}, \citet{libralato2026}, and \citet{callingham2022}, with the latter study also considering a tentative association ($p\sim 10\%$) with the $\lq$low-energy'-Kraken-Heracles progenitor \citep[LKH,][]{massari19,kruijssen2020,horta2021,massari2026}. However, based on our orbital analysis, the cluster orbital energy appears too high and the orbit too retrograde to support such a connection. 
%---------------------------------------------------------------
\section{Summary and conclusion}
\label{sec:conclusion}
%---------------------------------------------------------------

In this paper, we analysed low-resolution spectroscopic data obtained with MODS@LBT to derive the systemic heliocentric radial velocities and the metallicity for a set of faint and/or highly extinct stellar clusters targeted by the MGCS, whose high-resolution spectroscopic investigation remains challenging with current facilities.
For three systems, i.e. Koposov 1, Koposov 2, and Mu\~noz 1, we validated our measurements against existing spectroscopic determinations available in the literature \citep{munoz2012,geha2026} and consistently find excellent agreement. In addition, we present results for Pfleiderer 2 and RLGC2, providing their systemic velocities and spectroscopic metallicity for the first time. We find that Pfleiderer 2 has a $\vsys = 6\pm5$ km $\mathrm{s^{-1}}$ and $\feh=-0.75\pm0.09$ dex, while RLGC2 has $\vsys = -313\pm6$ km $\mathrm{s^{-1}}$ and $\feh=-2.33\pm0.13$ dex.

For four of the clusters, these new measurements have enabled the first accurate orbital analysis aimed at constraining their origin. We find that one cluster is likely dynamically associated with the Sagittarius dwarf spheroidal galaxy (Koposov 1), one is possibly linked to the \textit{Gaia}–Sausage–Enceladus merger (RLGC2), one is consistent with an in-situ origin moving on a resonant orbit (Pfleiderer 2), and one remains ungrouped (Koposov 2). Finally, we confirm previous findings that Mu\~noz 1 occupies a similar region of the integrals of motion space as other clusters associated with the Sag dSph \citep[e.g.][]{callingham2022}. Nevertheless, it is important to stress that dynamical information alone can sometimes be misleading as orbital parameters alone might not always accurately trace the origin of GCs \citep[see e.g.][]{pfeffer2020,pagnini2023,ceccarelli2025}. A definitive assessment of the nature and origin of these systems ultimately requires the combination of dynamics with detailed chemical abundances and precise age determinations.

In recent years the community has started an extraordinary effort towards a comprehensive census of the MW GC system, especially focusing on observationally challenging clusters. On the photometric side, the final opportunities provided by \textit{HST} before decommissioning are being leveraged through dedicated surveys such as the MGCS \citep{massari2025}. Parallel efforts are advancing on the chemical front, including studies based either on available data and current facilities \citep[see e.g.][]{geisler2021,pace2023,ferraro25,garro2026}, and next generation wide-field spectroscopic surveys, such as MOONS and 4MOST, with programmes specifically designed to study Galactic stellar clusters and dwarf MW satellites \citep{gonzalez2020,lucatello2023,skuladottir2023}.
Together, these complementary initiatives promise to complete our understanding of the formation history of the MW by providing a complete and homogeneous characterisation of its stellar cluster populations.

\begin{acknowledgements}

Based on data acquired using the Large Binocular Telescope (LBT) through the program IT-2024B-009 (P.I. E. Ceccarelli). The LBT is an international collaboration among institutions in the United States, Italy, and Germany. LBT Corporation partners are The University of Arizona on behalf of the Arizona university system; Istituto Nazionale di Astrofisica, Italy; LBT Beteiligungsgesellschaft, Germany, representing the Max-Planck Society, the Astrophysical Institute Potsdam, and Heidelberg University; The Ohio State University; and The Research Corporation, on behalf of The University of Notre Dame, University of Minnesota, and University of Virginia.

This work has made use of data from the European Space Agency (ESA) mission \textit{Gaia} \url{https://www.cosmos.esa.int/gaia}), processed by the \textit{Gaia} Data Processing and Analysis Consortium (DPAC, \url{https://www.cosmos.esa.int/web/gaia/dpac/consortium}). Funding for the DPAC has been provided by national institutions, in particular the institutions participating in the \textit{Gaia} Multilateral Agreement. 

DM acknowledges financial support from PRIN-MIUR-22 ``CHRONOS: adjusting the clock(s) to unveil the CHRONO-chemo-dynamical Structure of the Galaxy” (PI: S. Cassisi) granted by the European Union - Next Generation EU.    

EC, MB, AM, and MDL acknowledge financial support from the project ``LEGO – Reconstructing the building blocks of the Galaxy by chemical tagging” (PI: Mucciarelli) granted by the Italian MUR through contract PRIN2022LLP8TK\_001.

EC, MB, and AM are grateful to A. Gargiulo for her assistance with the reduction of MODS data using \texttt{SIPGI}.

EC thanks F. Cusano for his help in preparing MODS masks, and A. Della Croce for assistance with the figures.
    
\end{acknowledgements}

\bibliographystyle{aa}
\bibliography{MODS_MGCS}

@ARTICLE{libralato2026,
       author = {{Libralato}, M. and {Bellini}, A. and {Massari}, D. and {Bellazzini}, M. and {Aguado-Agelet}, F. and {Cassisi}, S. and {Ceccarelli}, E. and {Dalessandro}, E. and {Dodd}, E. and {Ferraro}, F.~R. and {Gallart}, C. and {Lanzoni}, B. and {Monelli}, M. and {Mucciarelli}, A. and {Pancino}, E. and {Pascale}, R. and {Rosignoli}, L. and {Salaris}, M. and {Saracino}, S. and {Zerbinati}, C.},
        title = "{The Hubble Missing Globular Cluster Survey: III. Astro-photometric catalogs, artificial-star tests, and improved absolute proper motions}",
      journal = {\aap},
         year = 2026,
        month = may,
       volume = {709},
          eid = {A140},
        pages = {A140},
          doi = {10.1051/0004-6361/202659363},
archivePrefix = {arXiv},
       eprint = {2603.25848},
 primaryClass = {astro-ph.GA},
       adsurl = {https://ui.adsabs.harvard.edu/abs/2026A&A...709A.140L}
}

@ARTICLE{gran2022,
       author = {{Gran}, F. and {Zoccali}, M. and {Saviane}, I. and {Valenti}, E. and {Rojas-Arriagada}, A. and {Contreras Ramos}, R. and {Hartke}, J. and {Carballo-Bello}, J.~A. and {Navarrete}, C. and {Rejkuba}, M. and {Olivares Carvajal}, J.},
        title = "{Hidden in the haystack: low-luminosity globular clusters towards the Milky Way bulge}",
      journal = {\mnras},
         year = 2022,
        month = feb,
       volume = {509},
       number = {4},
        pages = {4962-4981},
          doi = {10.1093/mnras/stab2463},
archivePrefix = {arXiv},
       eprint = {2108.11922},
 primaryClass = {astro-ph.GA},
       adsurl = {https://ui.adsabs.harvard.edu/abs/2022MNRAS.509.4962G}
}

@ARTICLE{belokurov2010,
       author = {{Belokurov}, V. and {Walker}, M.~G. and {Evans}, N.~W. and {Gilmore}, G. and {Irwin}, M.~J. and {Just}, D. and {Koposov}, S. and {Mateo}, M. and {Olszewski}, E. and {Watkins}, L. and {Wyrzykowski}, L.},
        title = "{Big Fish, Little Fish: Two New Ultra-faint Satellites of the Milky Way}",
      journal = {\apjl},
         year = 2010,
        month = mar,
       volume = {712},
       number = {1},
        pages = {L103-L106},
          doi = {10.1088/2041-8205/712/1/L103},
archivePrefix = {arXiv},
       eprint = {1002.0504},
 primaryClass = {astro-ph.GA},
       adsurl = {https://ui.adsabs.harvard.edu/abs/2010ApJ...712L.103B}
}

@ARTICLE{webb&carlberg2021,
       author = {{Webb}, Jeremy J. and {Carlberg}, Raymond G.},
        title = "{The likelihood of undiscovered globular clusters in the outskirts of the Milky Way}",
      journal = {\mnras},
         year = 2021,
        month = apr,
       volume = {502},
       number = {3},
        pages = {4547-4557},
          doi = {10.1093/mnras/stab353},
archivePrefix = {arXiv},
       eprint = {2012.08535},
 primaryClass = {astro-ph.GA},
       adsurl = {https://ui.adsabs.harvard.edu/abs/2021MNRAS.502.4547W}
}

@ARTICLE{newberg2009ApJ,
       author = {{Newberg}, Heidi Jo and {Yanny}, Brian and {Willett}, Benjamin A.},
        title = "{Discovery of a New, Polar-Orbiting Debris Stream in the Milky Way Stellar Halo}",
      journal = {\apjl},
         year = 2009,
        month = aug,
       volume = {700},
       number = {2},
        pages = {L61-L64},
          doi = {10.1088/0004-637X/700/2/L61},
archivePrefix = {arXiv},
       eprint = {0906.3291},
 primaryClass = {astro-ph.GA},
       adsurl = {https://ui.adsabs.harvard.edu/abs/2009ApJ...700L..61N}
}

@ARTICLE{massari2026,
       author = {{Massari}, Davide and {Zerbinati}, Chiara and {Fanelli}, Cristiano and {Helmi}, Amina and {Ceccarelli}, Edoardo and {Aguado-Agelet}, Fernando and {Cassisi}, Santi and {Wempe}, Ewoud and {Monelli}, Matteo and {Bellini}, Andrea and {Callingham}, Thomas and {Woudenberg}, Hanneke C. and {Cohen}, Roger and {Gallart}, Carme and {Pancino}, Elena and {Saracino}, Sara and {Salaris}, Maurizio and {Mucciarelli}, Alessio},
        title = "{Proof that the Milky Way experienced a significant merger only 1.5 billion years after the Big Bang}",
      journal = {arXiv e-prints},
         year = 2026,
        month = jan,
          eid = {arXiv:2601.18896},
        pages = {arXiv:2601.18896},
          doi = {10.48550/arXiv.2601.18896},
archivePrefix = {arXiv},
       eprint = {2601.18896},
 primaryClass = {astro-ph.GA},
       adsurl = {https://ui.adsabs.harvard.edu/abs/2026arXiv260118896M}
}

@ARTICLE{starkenburg2010,
       author = {{Starkenburg}, E. and {Hill}, V. and {Tolstoy}, E. and {Gonz{\'a}lez Hern{\'a}ndez}, J.~I. and {Irwin}, M. and {Helmi}, A. and {Battaglia}, G. and {Jablonka}, P. and {Tafelmeyer}, M. and {Shetrone}, M. and et al.},
        title = "{The NIR Ca ii triplet at low metallicity. Searching for extremely low-metallicity stars in classical dwarf galaxies}",
      journal = {\aap},
         year = 2010,
        month = apr,
       volume = {513},
          eid = {A34},
        pages = {A34},
          doi = {10.1051/0004-6361/200913759},
archivePrefix = {arXiv},
       eprint = {1002.2963},
 primaryClass = {astro-ph.SR},
       adsurl = {https://ui.adsabs.harvard.edu/abs/2010A&A...513A..34S}
}

@ARTICLE{carrera2013,
       author = {{Carrera}, R. and {Pancino}, E. and {Gallart}, C. and {del Pino}, A.},
        title = "{The near-infrared Ca II triplet as a metallicity indicator - II. Extension to extremely metal-poor metallicity regimes}",
      journal = {\mnras},
         year = 2013,
        month = sep,
       volume = {434},
       number = {2},
        pages = {1681-1691},
          doi = {10.1093/mnras/stt1126},
archivePrefix = {arXiv},
       eprint = {1306.3883},
 primaryClass = {astro-ph.GA},
       adsurl = {https://ui.adsabs.harvard.edu/abs/2013MNRAS.434.1681C}
}

@ARTICLE{navabi2026,
       author = {{Navabi}, M. and {Carrera}, R. and {No{\"e}l}, N.~E.~D. and {Gallart}, C. and {Pancino}, E. and {De Leo}, M.},
        title = "{Revisiting the near-infrared calcium triplet as metallicity indicator}",
      journal = {\mnras},
         year = 2026,
        month = feb,
       volume = {546},
       number = {2},
          eid = {stag019},
        pages = {stag019},
          doi = {10.1093/mnras/stag019},
archivePrefix = {arXiv},
       eprint = {2512.20574},
 primaryClass = {astro-ph.SR},
       adsurl = {https://ui.adsabs.harvard.edu/abs/2026MNRAS.546ag019N}
}

@ARTICLE{bellazzini2020,
       author = {{Bellazzini}, M. and {Ibata}, R. and {Malhan}, K. and {Martin}, N. and {Famaey}, B. and {Thomas}, G.},
        title = "{Globular clusters in the Sagittarius stream. Revising members and candidates with Gaia DR2}",
      journal = {\aap},
         year = 2020,
        month = apr,
       volume = {636},
          eid = {A107},
        pages = {A107},
          doi = {10.1051/0004-6361/202037621},
archivePrefix = {arXiv},
       eprint = {2003.07871},
 primaryClass = {astro-ph.GA},
       adsurl = {https://ui.adsabs.harvard.edu/abs/2020A&A...636A.107B}
}

@ARTICLE{ceccarelli2025,
       author = {{Ceccarelli}, E. and {Massari}, D. and {Aguado-Agelet}, F. and {Mucciarelli}, A. and {Cassisi}, S. and {Monelli}, M. and {Pancino}, E. and {Salaris}, M. and {Saracino}, S.},
        title = "{Cluster Ages to Reconstruct the Milky Way Assembly (CARMA): III. NGC 288 as the first Splashed globular cluster}",
      journal = {\aap},
         year = 2025,
        month = dec,
       volume = {704},
          eid = {A256},
        pages = {A256},
          doi = {10.1051/0004-6361/202554354},
archivePrefix = {arXiv},
       eprint = {2503.02939},
 primaryClass = {astro-ph.GA},
       adsurl = {https://ui.adsabs.harvard.edu/abs/2025A&A...704A.256C}
}

@ARTICLE{harris2010,
       author = {{Harris}, William E.},
        title = "{A New Catalog of Globular Clusters in the Milky Way}",
      journal = {arXiv e-prints},
         year = 2010,
        month = dec,
          eid = {arXiv:1012.3224},
        pages = {arXiv:1012.3224},
          doi = {10.48550/arXiv.1012.3224},
archivePrefix = {arXiv},
       eprint = {1012.3224},
 primaryClass = {astro-ph.GA},
       adsurl = {https://ui.adsabs.harvard.edu/abs/2010arXiv1012.3224H}
}

@ARTICLE{pagnini2023,
       author = {{Pagnini}, G. and {Di Matteo}, P. and {Khoperskov}, S. and {Mastrobuono-Battisti}, A. and {Haywood}, M. and {Renaud}, F. and {Combes}, F.},
        title = "{The distribution of globular clusters in kinematic spaces does not trace the accretion history of the host galaxy}",
      journal = {\aap},
         year = 2023,
        month = may,
       volume = {673},
          eid = {A86},
        pages = {A86},
          doi = {10.1051/0004-6361/202245128},
archivePrefix = {arXiv},
       eprint = {2210.04245},
 primaryClass = {astro-ph.GA},
       adsurl = {https://ui.adsabs.harvard.edu/abs/2023A&A...673A..86P}
}

@ARTICLE{koposov2017,
       author = {{Koposov}, Sergey E. and {Belokurov}, V. and {Torrealba}, G.},
        title = "{Gaia 1 and 2. A pair of new Galactic star clusters}",
      journal = {\mnras},
         year = 2017,
        month = sep,
       volume = {470},
       number = {3},
        pages = {2702-2709},
          doi = {10.1093/mnras/stx1182},
archivePrefix = {arXiv},
       eprint = {1702.01122},
 primaryClass = {astro-ph.GA},
       adsurl = {https://ui.adsabs.harvard.edu/abs/2017MNRAS.470.2702K}
}

@ARTICLE{fadely2011,
       author = {{Fadely}, Ross and {Willman}, Beth and {Geha}, Marla and {Walsh}, Shane and {Mu{\~n}oz}, Ricardo R. and {Jerjen}, Helmut and {Vargas}, Luis C. and {Da Costa}, Gary S.},
        title = "{Segue 3: An Old, Extremely Low Luminosity Star Cluster in the Milky Way's Halo}",
      journal = {\aj},
         year = 2011,
        month = sep,
       volume = {142},
       number = {3},
          eid = {88},
        pages = {88},
          doi = {10.1088/0004-6256/142/3/88},
archivePrefix = {arXiv},
       eprint = {1107.3151},
 primaryClass = {astro-ph.GA},
       adsurl = {https://ui.adsabs.harvard.edu/abs/2011AJ....142...88F}
}

@ARTICLE{sitnova2024,
       author = {{Sitnova}, T.~M. and {Yuan}, Z. and {Matsuno}, T. and {Mashonkina}, L.~I. and {Alexeeva}, S.~A. and {Holmbeck}, E. and {Sestito}, F. and {Lombardo}, L. and {Banerjee}, P. and {Martin}, N.~F. and et al.},
        title = "{HR-GO: I. Comprehensive NLTE abundance analysis of the Cetus stream}",
      journal = {\aap},
         year = 2024,
        month = oct,
       volume = {690},
          eid = {A331},
        pages = {A331},
          doi = {10.1051/0004-6361/202450981},
archivePrefix = {arXiv},
       eprint = {2408.16107},
 primaryClass = {astro-ph.GA},
       adsurl = {https://ui.adsabs.harvard.edu/abs/2024A&A...690A.331S}
}

@ARTICLE{skuladottir2023,
       author = {{Sk{\'u}lad{\'o}ttir}, {\'A}. and {Puls}, A.~A. and {Amarsi}, A.~M. and {Battaglia}, G. and {Buder}, S. and {Campbell}, S. and {Cardona-Barrero}, S. and {Christlieb}, N. and {Feuillet}, D.~K. and {Gelli}, V. and et al.},
        title = "{The 4MOST Survey of Dwarf Galaxies and their Stellar Streams (4DWARFS)}",
      journal = {The Messenger},
         year = 2023,
        month = mar,
       volume = {190},
        pages = {19-21},
          doi = {10.18727/0722-6691/5304},
       adsurl = {https://ui.adsabs.harvard.edu/abs/2023Msngr.190...19S}
}

@ARTICLE{massari25,
       author = {{Massari}, Davide},
        title = "{Origin of the System of Globular Clusters in the Milky Way{\textemdash}Gaia eDR3 Edition}",
      journal = {Research Notes of the American Astronomical Society},
         year = 2025,
        month = mar,
       volume = {9},
       number = {3},
          eid = {64},
        pages = {64},
          doi = {10.3847/2515-5172/adc375},
archivePrefix = {arXiv},
       eprint = {2503.14657},
 primaryClass = {astro-ph.GA},
       adsurl = {https://ui.adsabs.harvard.edu/abs/2025RNAAS...9...64M}
}

@ARTICLE{mau2020,
       author = {{Mau}, S. and {Cerny}, W. and {Pace}, A.~B. and {Choi}, Y. and {Drlica-Wagner}, A. and {Santana-Silva}, L. and {Riley}, A.~H. and {Erkal}, D. and {Stringfellow}, G.~S. and {Adam{\'o}w}, M. and et al.},
        title = "{Two Ultra-faint Milky Way Stellar Systems Discovered in Early Data from the DECam Local Volume Exploration Survey}",
      journal = {\apj},
         year = 2020,
        month = feb,
       volume = {890},
       number = {2},
          eid = {136},
        pages = {136},
          doi = {10.3847/1538-4357/ab6c67},
archivePrefix = {arXiv},
       eprint = {1912.03301},
 primaryClass = {astro-ph.GA},
       adsurl = {https://ui.adsabs.harvard.edu/abs/2020ApJ...890..136M}
}

@ARTICLE{conn2018,
       author = {{Conn}, Blair C. and {Jerjen}, Helmut and {Kim}, Dongwon and {Schirmer}, Mischa},
        title = "{On the Nature of Ultra-faint Dwarf Galaxy Candidates. I. DES1, Eridanus III, and Tucana V}",
      journal = {\apj},
         year = 2018,
        month = jan,
       volume = {852},
       number = {2},
          eid = {68},
        pages = {68},
          doi = {10.3847/1538-4357/aa9eda},
archivePrefix = {arXiv},
       eprint = {1712.01439},
 primaryClass = {astro-ph.GA},
       adsurl = {https://ui.adsabs.harvard.edu/abs/2018ApJ...852...68C}
}

@ARTICLE{mucciarelli2017,
       author = {{Mucciarelli}, A. and {Bellazzini}, M. and {Ibata}, R. and {Romano}, D. and {Chapman}, S.~C. and {Monaco}, L.},
        title = "{Chemical abundances in the nucleus of the Sagittarius dwarf spheroidal galaxy}",
      journal = {\aap},
         year = 2017,
        month = sep,
       volume = {605},
          eid = {A46},
        pages = {A46},
          doi = {10.1051/0004-6361/201730707},
archivePrefix = {arXiv},
       eprint = {1705.03251},
 primaryClass = {astro-ph.GA},
       adsurl = {https://ui.adsabs.harvard.edu/abs/2017A&A...605A..46M}
}

@ARTICLE{minelli23,
       author = {{Minelli}, Alice and {Bellazzini}, Michele and {Mucciarelli}, Alessio and {Bonifacio}, Piercarlo and {Ibata}, Rodrigo and {Romano}, Donatella and {Monaco}, Lorenzo and {Caffau}, Elisabetta and {Dalessandro}, Emanuele and {Pascale}, Raffaele},
        title = "{The metallicity distribution in the core of the Sagittarius dwarf spheroidal: Minimising the metallicity biases}",
      journal = {\aap},
         year = 2023,
        month = jan,
       volume = {669},
          eid = {A54},
        pages = {A54},
          doi = {10.1051/0004-6361/202244890},
archivePrefix = {arXiv},
       eprint = {2211.06727},
 primaryClass = {astro-ph.GA},
       adsurl = {https://ui.adsabs.harvard.edu/abs/2023A&A...669A..54M}
}

@ARTICLE{robin2003,
       author = {{Robin}, A.~C. and {Reyl{\'e}}, C. and {Derri{\`e}re}, S. and {Picaud}, S.},
        title = "{A synthetic view on structure and evolution of the Milky Way}",
      journal = {\aap},
         year = 2003,
        month = oct,
       volume = {409},
        pages = {523-540},
          doi = {10.1051/0004-6361:20031117},
       adsurl = {https://ui.adsabs.harvard.edu/abs/2003A&A...409..523R}
}

@ARTICLE{cerny26,
       author = {{Cerny}, William and {Li}, Ting S. and {Pace}, Andrew B. and {Simon}, Joshua D. and {Geha}, Marla and {Ji}, Alexander P. and {Drlica-Wagner}, Alex and {Bruce}, Jordan and {Gnedin}, Oleg Y. and {Bell}, Eric F. and et al.},
        title = "{A Chemodynamical Census of the Milky Way's Ultra-Faint Compact Satellites. I. A First Population-Level Look at the Internal Kinematics and Metallicities of 19 Extremely-Low-Mass Halo Stellar Systems}",
      journal = {arXiv e-prints},
         year = 2026,
        month = feb,
          eid = {arXiv:2602.17652},
        pages = {arXiv:2602.17652},
archivePrefix = {arXiv},
       eprint = {2602.17652},
 primaryClass = {astro-ph.GA},
       adsurl = {https://ui.adsabs.harvard.edu/abs/2026arXiv260217652C}
}

@ARTICLE{laevens2015,
       author = {{Laevens}, Benjamin P.~M. and {Martin}, Nicolas F. and {Bernard}, Edouard J. and {Schlafly}, Edward F. and {Sesar}, Branimir and {Rix}, Hans-Walter and {Bell}, Eric F. and {Ferguson}, Annette M.~N. and {Slater}, Colin T. and {Sweeney}, William E. and et al.},
        title = "{Sagittarius II, Draco II and Laevens 3: Three New Milky Way Satellites Discovered in the Pan-STARRS 1 3{\ensuremath{\pi}} Survey}",
      journal = {\apj},
         year = 2015,
        month = nov,
       volume = {813},
       number = {1},
          eid = {44},
        pages = {44},
          doi = {10.1088/0004-637X/813/1/44},
archivePrefix = {arXiv},
       eprint = {1507.07564},
 primaryClass = {astro-ph.GA},
       adsurl = {https://ui.adsabs.harvard.edu/abs/2015ApJ...813...44L}
}

@ARTICLE{lucatello2023,
       author = {{Lucatello}, S. and {Bragaglia}, A. and {Vallenari}, A. and {Cantat-Gaudin}, T. and {Kuzma}, P. and {Guarcello}, M.~G. and {Spina}, L. and {Aguado}, D. and {Carrera}, R. and {Castro-Ginard}, A. and et al.},
        title = "{Stellar Clusters in 4MOST}",
      journal = {The Messenger},
         year = 2023,
        month = mar,
       volume = {190},
        pages = {13-16},
          doi = {10.18727/0722-6691/5302},
       adsurl = {https://ui.adsabs.harvard.edu/abs/2023Msngr.190...13L}
}

@ARTICLE{ferraro25,
       author = {{Ferraro}, F.~R. and {Chiappino}, L. and {Bartolomei}, A. and {Origlia}, L. and {Fanelli}, C. and {Lanzoni}, B. and {Pallanca}, C. and {Loriga}, M. and {Leanza}, S. and {Valenti}, E. and {Romano}, D. and {Mucciarelli}, A. and {Massari}, D. and {Cadelano}, M. and {Dalessandro}, E. and {Crociati}, C. and {Rich}, R.~M.},
        title = "{The Bulge Cluster Origin (BulCO) survey at the ESO-VLT: Probing the early history of the Milky Way assembly. Design and first results in Liller 1}",
      journal = {\aap},
         year = 2025,
        month = apr,
       volume = {696},
          eid = {A179},
        pages = {A179},
          doi = {10.1051/0004-6361/202554092},
archivePrefix = {arXiv},
       eprint = {2503.14642},
 primaryClass = {astro-ph.GA},
       adsurl = {https://ui.adsabs.harvard.edu/abs/2025A&A...696A.179F}
}

@ARTICLE{geisler2021,
       author = {{Geisler}, D. and {Villanova}, S. and {O'Connell}, J.~E. and {Cohen}, R.~E. and {Moni Bidin}, C. and {Fern{\'a}ndez-Trincado}, J.~G. and {Mu{\~n}oz}, C. and {Minniti}, D. and {Zoccali}, M. and {Rojas-Arriagada}, A. and {Contreras Ramos}, R. and {Catelan}, M. and {Mauro}, F. and {Cort{\'e}s}, C. and {Ferreira Lopes}, C.~E. and {Arentsen}, A. and {Starkenburg}, E. and {Martin}, N.~F. and {Tang}, B. and {Parisi}, C. and {Alonso-Garc{\'\i}a}, J. and {Gran}, F. and {Cunha}, K. and {Smith}, V. and {Majewski}, S.~R. and {J{\"o}nsson}, H. and {Garc{\'\i}a-Hern{\'a}ndez}, D.~A. and {Horta}, D. and {M{\'e}sz{\'a}ros}, S. and {Monaco}, L. and {Monachesi}, A. and {Mu{\~n}oz}, R.~R. and {Brownstein}, J. and {Beers}, T.~C. and {Lane}, R.~R. and {Barbuy}, B. and {Sobeck}, J. and {Henao}, L. and {Gonz{\'a}lez-D{\'\i}az}, D. and {Miranda}, R.~E. and {Reinarz}, Y. and {Santander}, T.~A.},
        title = "{CAPOS: The bulge Cluster APOgee Survey. I. Overview and initial ASPCAP results}",
      journal = {\aap},
         year = 2021,
        month = aug,
       volume = {652},
          eid = {A157},
        pages = {A157},
          doi = {10.1051/0004-6361/202140436},
archivePrefix = {arXiv},
       eprint = {2106.00024},
 primaryClass = {astro-ph.GA},
       adsurl = {https://ui.adsabs.harvard.edu/abs/2021A&A...652A.157G}
}

@ARTICLE{pfeffer2020,
       author = {{Pfeffer}, Joel L. and {Trujillo-Gomez}, Sebastian and {Kruijssen}, J.~M.~D. and {Crain}, Robert A. and {Hughes}, Meghan E. and {Reina-Campos}, Marta and {Bastian}, Nate},
        title = "{Predicting accreted satellite galaxy masses and accretion redshifts based on globular cluster orbits in the E-MOSAICS simulations}",
      journal = {\mnras},
         year = 2020,
        month = dec,
       volume = {499},
       number = {4},
        pages = {4863-4875},
          doi = {10.1093/mnras/staa3109},
archivePrefix = {arXiv},
       eprint = {2003.00076},
 primaryClass = {astro-ph.GA},
       adsurl = {https://ui.adsabs.harvard.edu/abs/2020MNRAS.499.4863P}
}

@ARTICLE{hayes2020,
       author = {{Hayes}, Christian R. and {Majewski}, Steven R. and {Hasselquist}, Sten and {Anguiano}, Borja and {Shetrone}, Matthew and {Law}, David R. and {Schiavon}, Ricardo P. and {Cunha}, Katia and {Smith}, Verne V. and {Beaton}, Rachael L. and {Price-Whelan}, Adrian M. and {Allende Prieto}, Carlos and {Battaglia}, Giuseppina and {Bizyaev}, Dmitry and {Brownstein}, Joel R. and {Cohen}, Roger E. and {Frinchaboy}, Peter M. and {Garc{\'\i}a-Hern{\'a}ndez}, D.~A. and {Lacerna}, Ivan and {Lane}, Richard R. and {M{\'e}sz{\'a}ros}, Szabolcs and {Bidin}, Christian Moni and {M{\~{u}}noz}, Ricardo R. and {Nidever}, David L. and {Oravetz}, Audrey and {Oravetz}, Daniel and {Pan}, Kaike and {Roman-Lopes}, Alexandre and {Sobeck}, Jennifer and {Stringfellow}, Guy},
        title = "{Metallicity and {\ensuremath{\alpha}}-Element Abundance Gradients along the Sagittarius Stream as Seen by APOGEE}",
      journal = {\apj},
         year = 2020,
        month = jan,
       volume = {889},
       number = {1},
          eid = {63},
        pages = {63},
          doi = {10.3847/1538-4357/ab62ad},
archivePrefix = {arXiv},
       eprint = {1912.06707},
 primaryClass = {astro-ph.GA},
       adsurl = {https://ui.adsabs.harvard.edu/abs/2020ApJ...889...63H}
}

@ARTICLE{taylor2025,
       author = {{Taylor}, Ethan D. and {Read}, Justin I. and {Orkney}, Matthew D.~A. and {Kim}, Stacy Y. and {Pontzen}, Andrew and {Agertz}, Oscar and {Rey}, Martin P. and {Andersson}, Eric P. and {Collins}, Michelle L.~M. and {Yates}, Robert M.},
        title = "{The emergence of globular clusters and globular-cluster-like dwarfs}",
      journal = {\nat},
         year = 2025,
        month = sep,
       volume = {645},
       number = {8080},
        pages = {327-331},
          doi = {10.1038/s41586-025-09494-x},
archivePrefix = {arXiv},
       eprint = {2509.09582},
 primaryClass = {astro-ph.GA},
       adsurl = {https://ui.adsabs.harvard.edu/abs/2025Natur.645..327T}
}

@ARTICLE{munoz2018,
       author = {{Mu{\~n}oz}, Ricardo R. and {C{\^o}t{\'e}}, Patrick and {Santana}, Felipe A. and {Geha}, Marla and {Simon}, Joshua D. and {Oyarz{\'u}n}, Grecco A. and {Stetson}, Peter B. and {Djorgovski}, S.~G.},
        title = "{A MegaCam Survey of Outer Halo Satellites. III. Photometric and Structural Parameters}",
      journal = {\apj},
         year = 2018,
        month = jun,
       volume = {860},
       number = {1},
          eid = {66},
        pages = {66},
          doi = {10.3847/1538-4357/aac16b},
archivePrefix = {arXiv},
       eprint = {1806.06891},
 primaryClass = {astro-ph.GA},
       adsurl = {https://ui.adsabs.harvard.edu/abs/2018ApJ...860...66M}
}

@ARTICLE{geha2026,
       author = {{Geha}, Marla and {Pelliccia}, Debora and {Prochaska}, J. Xavier and {Cerny}, William and {Davies}, Frederick B. and {Hennawi}, Joseph and {Holden}, Brad and {Reichwein}, Dusty and {Westfall}, Kyle B.},
        title = "{The Keck/DEIMOS Stellar Archive. I. Uniform Velocities and Metallicities for 78 Milky Way Dwarf Galaxies and Globular Clusters}",
      journal = {\apj},
         year = 2026,
        month = mar,
       volume = {999},
       number = {1},
          eid = {140},
        pages = {140},
          doi = {10.3847/1538-4357/ae290d},
archivePrefix = {arXiv},
       eprint = {2602.10200},
 primaryClass = {astro-ph.GA},
       adsurl = {https://ui.adsabs.harvard.edu/abs/2026ApJ...999..140G}
}

@ARTICLE{belokurov2006,
       author = {{Belokurov}, V. and {Zucker}, D.~B. and {Evans}, N.~W. and {Gilmore}, G. and {Vidrih}, S. and {Bramich}, D.~M. and {Newberg}, H.~J. and {Wyse}, R.~F.~G. and {Irwin}, M.~J. and {Fellhauer}, M. and {Hewett}, P.~C. and {Walton}, N.~A. and {Wilkinson}, M.~I. and {Cole}, N. and {Yanny}, B. and {Rockosi}, C.~M. and {Beers}, T.~C. and {Bell}, E.~F. and {Brinkmann}, J. and {Ivezi{\'c}}, {\v{Z}}. and {Lupton}, R.},
        title = "{The Field of Streams: Sagittarius and Its Siblings}",
      journal = {\apjl},
         year = 2006,
        month = may,
       volume = {642},
       number = {2},
        pages = {L137-L140},
          doi = {10.1086/504797},
archivePrefix = {arXiv},
       eprint = {astro-ph/0605025},
 primaryClass = {astro-ph},
       adsurl = {https://ui.adsabs.harvard.edu/abs/2006ApJ...642L.137B}
}

@ARTICLE{carrera2007,
       author = {{Carrera}, R. and {Gallart}, C. and {Pancino}, E. and {Zinn}, R.},
        title = "{The Infrared Ca II Triplet as Metallicity Indicator}",
      journal = {\aj},
         year = 2007,
        month = sep,
       volume = {134},
       number = {3},
        pages = {1298},
          doi = {10.1086/520803},
archivePrefix = {arXiv},
       eprint = {0705.3335},
 primaryClass = {astro-ph},
       adsurl = {https://ui.adsabs.harvard.edu/abs/2007AJ....134.1298C}
}

@ARTICLE{kim2016,
       author = {{Kim}, Dongwon and {Jerjen}, Helmut and {Mackey}, Dougal and {Da Costa}, Gary S. and {Milone}, Antonino P.},
        title = "{KIM 3: An Ultra-faint Star Cluster in the Constellation of Centaurus}",
      journal = {\apj},
         year = 2016,
        month = apr,
       volume = {820},
       number = {2},
          eid = {119},
        pages = {119},
          doi = {10.3847/0004-637X/820/2/119},
archivePrefix = {arXiv},
       eprint = {1512.03530},
 primaryClass = {astro-ph.GA},
       adsurl = {https://ui.adsabs.harvard.edu/abs/2016ApJ...820..119K}
}

@ARTICLE{garro2020,
       author = {{Garro}, E.~R. and {Minniti}, D. and {G{\'o}mez}, M. and {Alonso-Garc{\'\i}a}, J. and {Barb{\'a}}, R.~H. and {Barbuy}, B. and {Clari{\'a}}, J.~J. and {Chen{\'e}}, A.~N. and {Dias}, B. and {Hempel}, M. and {Ivanov}, V.~D. and {Lucas}, P.~W. and {Majaess}, D. and {Mauro}, F. and {Moni Bidin}, C. and {Palma}, T. and {Pullen}, J.~B. and {Saito}, R.~K. and {Smith}, L. and {Surot}, F. and {Ram{\'\i}rez Alegr{\'\i}a}, S. and {Rejkuba}, M. and {Ripepi}, V. and {Fern{\'a}ndez Trincado}, J.},
        title = "{VVVX-Gaia discovery of a low luminosity globular cluster in the Milky Way disk}",
      journal = {\aap},
         year = 2020,
        month = oct,
       volume = {642},
          eid = {L19},
        pages = {L19},
          doi = {10.1051/0004-6361/202039233},
archivePrefix = {arXiv},
       eprint = {2010.02113},
 primaryClass = {astro-ph.GA},
       adsurl = {https://ui.adsabs.harvard.edu/abs/2020A&A...642L..19G}
}

@ARTICLE{minniti2011,
       author = {{Minniti}, D. and {Hempel}, M. and {Toledo}, I. and {Ivanov}, V.~D. and {Alonso-Garc{\'\i}a}, J. and {Saito}, R.~K. and {Catelan}, M. and {Geisler}, D. and {Jord{\'a}n}, A. and {Borissova}, J. and {Zoccali}, M. and {Kurtev}, R. and {Carraro}, G. and {Barbuy}, B. and {Clari{\'a}}, J. and {Rejkuba}, M. and {Emerson}, J. and {Moni Bidin}, C.},
        title = "{Discovery of VVV CL001. A low-mass globular cluster next to UKS 1 in the direction of the Galactic bulge}",
      journal = {\aap},
         year = 2011,
        month = mar,
       volume = {527},
          eid = {A81},
        pages = {A81},
          doi = {10.1051/0004-6361/201015795},
archivePrefix = {arXiv},
       eprint = {1012.2450},
 primaryClass = {astro-ph.GA},
       adsurl = {https://ui.adsabs.harvard.edu/abs/2011A&A...527A..81M}
}

@ARTICLE{ryu2018,
       author = {{Ryu}, Jinhyuk and {Lee}, Myung Gyoon},
        title = "{Discovery of Two New Globular Clusters in the Milky Way}",
      journal = {\apjl},
         year = 2018,
        month = aug,
       volume = {863},
       number = {2},
          eid = {L38},
        pages = {L38},
          doi = {10.3847/2041-8213/aad8b7},
archivePrefix = {arXiv},
       eprint = {1808.03455},
 primaryClass = {astro-ph.GA},
       adsurl = {https://ui.adsabs.harvard.edu/abs/2018ApJ...863L..38R}
}

@ARTICLE{ortolani2009,
       author = {{Ortolani}, S. and {Bonatto}, C. and {Bica}, E. and {Barbuy}, B.},
        title = "{Pfleiderer 2: Identification of A New Globular Cluster in the Galaxy}",
      journal = {\aj},
         year = 2009,
        month = sep,
       volume = {138},
       number = {3},
        pages = {889-894},
          doi = {10.1088/0004-6256/138/3/889},
archivePrefix = {arXiv},
       eprint = {0907.1225},
 primaryClass = {astro-ph.GA},
       adsurl = {https://ui.adsabs.harvard.edu/abs/2009AJ....138..889O}
}

@ARTICLE{munoz2012,
       author = {{Mu{\~n}oz}, R.~R. and {Geha}, M. and {C{\^o}t{\'e}}, P. and {Vargas}, L.~C. and {Santana}, F.~A. and {Stetson}, P. and {Simon}, J.~D. and {Djorgovski}, S.~G.},
        title = "{The Discovery of an Ultra-faint Star Cluster in the Constellation of Ursa Minor}",
      journal = {\apjl},
         year = 2012,
        month = jul,
       volume = {753},
       number = {1},
          eid = {L15},
        pages = {L15},
          doi = {10.1088/2041-8205/753/1/L15},
archivePrefix = {arXiv},
       eprint = {1204.5750},
 primaryClass = {astro-ph.GA},
       adsurl = {https://ui.adsabs.harvard.edu/abs/2012ApJ...753L..15M}
}

@ARTICLE{GC21,
       author = {{Gaia Collaboration} and {Brown}, A.~G.~A. and {Vallenari}, A. and {Prusti}, T. and {de Bruijne}, J.~H.~J. and {Babusiaux}, C. and {Biermann}, M. and {Creevey}, O.~L. and {Evans}, D.~W. and {Eyer}, L. and {Hutton}, A. and {Jansen}, F. and {Jordi}, C. and {Klioner}, S.~A. and {Lammers}, U. and {Lindegren}, L. and {Luri}, X. and {Mignard}, F. and {Panem}, C. and {Pourbaix}, D. and {Randich}, S. and {Sartoretti}, P. and {Soubiran}, C. and {Walton}, N.~A. and {Arenou}, F. and {Bailer-Jones}, C.~A.~L. and {Bastian}, U. and {Cropper}, M. and {Drimmel}, R. and {Katz}, D. and {Lattanzi}, M.~G. and {van Leeuwen}, F. and {Bakker}, J. and {Cacciari}, C. and {Casta{\~n}eda}, J. and {De Angeli}, F. and {Ducourant}, C. and {Fabricius}, C. and {Fouesneau}, M. and {Fr{\'e}mat}, Y. and {Guerra}, R. and {Guerrier}, A. and {Guiraud}, J. and {Jean-Antoine Piccolo}, A. and {Masana}, E. and {Messineo}, R. and {Mowlavi}, N. and {Nicolas}, C. and {Nienartowicz}, K. and {Pailler}, F. and {Panuzzo}, P. and {Riclet}, F. and {Roux}, W. and {Seabroke}, G.~M. and {Sordo}, R. and {Tanga}, P. and {Th{\'e}venin}, F. and {Gracia-Abril}, G. and {Portell}, J. and {Teyssier}, D. and {Altmann}, M. and {Andrae}, R. and {Bellas-Velidis}, I. and {Benson}, K. and {Berthier}, J. and {Blomme}, R. and {Brugaletta}, E. and {Burgess}, P.~W. and {Busso}, G. and {Carry}, B. and {Cellino}, A. and {Cheek}, N. and {Clementini}, G. and {Damerdji}, Y. and {Davidson}, M. and {Delchambre}, L. and {Dell'Oro}, A. and {Fern{\'a}ndez-Hern{\'a}ndez}, J. and {Galluccio}, L. and {Garc{\'\i}a-Lario}, P. and {Garcia-Reinaldos}, M. and {Gonz{\'a}lez-N{\'u}{\~n}ez}, J. and {Gosset}, E. and {Haigron}, R. and {Halbwachs}, J. -L. and {Hambly}, N.~C. and {Harrison}, D.~L. and {Hatzidimitriou}, D. and {Heiter}, U. and {Hern{\'a}ndez}, J. and {Hestroffer}, D. and {Hodgkin}, S.~T. and {Holl}, B. and {Jan{\ss}en}, K. and {Jevardat de Fombelle}, G. and {Jordan}, S. and {Krone-Martins}, A. and {Lanzafame}, A.~C. and {L{\"o}ffler}, W. and {Lorca}, A. and {Manteiga}, M. and {Marchal}, O. and {Marrese}, P.~M. and {Moitinho}, A. and {Mora}, A. and {Muinonen}, K. and {Osborne}, P. and {Pancino}, E. and {Pauwels}, T. and {Petit}, J. -M. and {Recio-Blanco}, A. and {Richards}, P.~J. and {Riello}, M. and {Rimoldini}, L. and {Robin}, A.~C. and {Roegiers}, T. and {Rybizki}, J. and {Sarro}, L.~M. and {Siopis}, C. and {Smith}, M. and {Sozzetti}, A. and {Ulla}, A. and {Utrilla}, E. and {van Leeuwen}, M. and {van Reeven}, W. and {Abbas}, U. and {Abreu Aramburu}, A. and {Accart}, S. and {Aerts}, C. and {Aguado}, J.~J. and {Ajaj}, M. and {Altavilla}, G. and {{\'A}lvarez}, M.~A. and {{\'A}lvarez Cid-Fuentes}, J. and {Alves}, J. and {Anderson}, R.~I. and {Anglada Varela}, E. and {Antoja}, T. and {Audard}, M. and {Baines}, D. and {Baker}, S.~G. and {Balaguer-N{\'u}{\~n}ez}, L. and {Balbinot}, E. and {Balog}, Z. and {Barache}, C. and {Barbato}, D. and {Barros}, M. and {Barstow}, M.~A. and {Bartolom{\'e}}, S. and {Bassilana}, J. -L. and {Bauchet}, N. and {Baudesson-Stella}, A. and {Becciani}, U. and {Bellazzini}, M. and {Bernet}, M. and {Bertone}, S. and {Bianchi}, L. and {Blanco-Cuaresma}, S. and {Boch}, T. and {Bombrun}, A. and {Bossini}, D. and {Bouquillon}, S. and {Bragaglia}, A. and {Bramante}, L. and {Breedt}, E. and {Bressan}, A. and {Brouillet}, N. and {Bucciarelli}, B. and {Burlacu}, A. and {Busonero}, D. and {Butkevich}, A.~G. and {Buzzi}, R. and {Caffau}, E. and {Cancelliere}, R. and {C{\'a}novas}, H. and {Cantat-Gaudin}, T. and {Carballo}, R. and {Carlucci}, T. and {Carnerero}, M.~I. and {Carrasco}, J.~M. and {Casamiquela}, L. and {Castellani}, M. and {Castro-Ginard}, A. and {Castro Sampol}, P. and {Chaoul}, L. and {Charlot}, P. and {Chemin}, L. and {Chiavassa}, A. and {Cioni}, M. -R.~L. and {Comoretto}, G. and {Cooper}, W.~J. and {Cornez}, T. and {Cowell}, S. and {Crifo}, F. and {Crosta}, M. and {Crowley}, C. and {Dafonte}, C. and {Dapergolas}, A. and {David}, M. and {David}, P. and {de Laverny}, P. and {De Luise}, F. and {De March}, R. and {De Ridder}, J. and {de Souza}, R. and {de Teodoro}, P. and {de Torres}, A. and {del Peloso}, E.~F. and {del Pozo}, E. and {Delbo}, M. and {Delgado}, A. and {Delgado}, H.~E. and {Delisle}, J. -B. and {Di Matteo}, P. and {Diakite}, S. and {Diener}, C. and {Distefano}, E. and {Dolding}, C. and {Eappachen}, D. and {Edvardsson}, B. and {Enke}, H. and {Esquej}, P. and {Fabre}, C. and {Fabrizio}, M. and {Faigler}, S. and {Fedorets}, G. and {Fernique}, P. and {Fienga}, A. and {Figueras}, F. and {Fouron}, C. and {Fragkoudi}, F. and {Fraile}, E. and {Franke}, F. and {Gai}, M. and {Garabato}, D. and {Garcia-Gutierrez}, A. and {Garc{\'\i}a-Torres}, M. and {Garofalo}, A. and {Gavras}, P. and {Gerlach}, E. and {Geyer}, R. and {Giacobbe}, P. and {Gilmore}, G. and {Girona}, S. and {Giuffrida}, G. and {Gomel}, R. and {Gomez}, A. and {Gonzalez-Santamaria}, I. and {Gonz{\'a}lez-Vidal}, J.~J. and {Granvik}, M. and {Guti{\'e}rrez-S{\'a}nchez}, R. and {Guy}, L.~P. and {Hauser}, M. and {Haywood}, M. and {Helmi}, A. and {Hidalgo}, S.~L. and {Hilger}, T. and {H{\l}adczuk}, N. and {Hobbs}, D. and {Holland}, G. and {Huckle}, H.~E. and {Jasniewicz}, G. and {Jonker}, P.~G. and {Juaristi Campillo}, J. and {Julbe}, F. and {Karbevska}, L. and {Kervella}, P. and {Khanna}, S. and {Kochoska}, A. and {Kontizas}, M. and {Kordopatis}, G. and {Korn}, A.~J. and {Kostrzewa-Rutkowska}, Z. and {Kruszy{\'n}ska}, K. and {Lambert}, S. and {Lanza}, A.~F. and {Lasne}, Y. and {Le Campion}, J. -F. and {Le Fustec}, Y. and {Lebreton}, Y. and {Lebzelter}, T. and {Leccia}, S. and {Leclerc}, N. and {Lecoeur-Taibi}, I. and {Liao}, S. and {Licata}, E. and {Lindstr{\o}m}, E.~P. and {Lister}, T.~A. and {Livanou}, E. and {Lobel}, A. and {Madrero Pardo}, P. and {Managau}, S. and {Mann}, R.~G. and {Marchant}, J.~M. and {Marconi}, M. and {Marcos Santos}, M.~M.~S. and {Marinoni}, S. and {Marocco}, F. and {Marshall}, D.~J. and {Martin Polo}, L. and {Mart{\'\i}n-Fleitas}, J.~M. and {Masip}, A. and {Massari}, D. and {Mastrobuono-Battisti}, A. and {Mazeh}, T. and {McMillan}, P.~J. and {Messina}, S. and {Michalik}, D. and {Millar}, N.~R. and {Mints}, A. and {Molina}, D. and {Molinaro}, R. and {Moln{\'a}r}, L. and {Montegriffo}, P. and {Mor}, R. and {Morbidelli}, R. and {Morel}, T. and {Morris}, D. and {Mulone}, A.~F. and {Munoz}, D. and {Muraveva}, T. and {Murphy}, C.~P. and {Musella}, I. and {Noval}, L. and {Ord{\'e}novic}, C. and {Orr{\`u}}, G. and {Osinde}, J. and {Pagani}, C. and {Pagano}, I. and {Palaversa}, L. and {Palicio}, P.~A. and {Panahi}, A. and {Pawlak}, M. and {Pe{\~n}alosa Esteller}, X. and {Penttil{\"a}}, A. and {Piersimoni}, A.~M. and {Pineau}, F. -X. and {Plachy}, E. and {Plum}, G. and {Poggio}, E. and {Poretti}, E. and {Poujoulet}, E. and {Pr{\v{s}}a}, A. and {Pulone}, L. and {Racero}, E. and {Ragaini}, S. and {Rainer}, M. and {Raiteri}, C.~M. and {Rambaux}, N. and {Ramos}, P. and {Ramos-Lerate}, M. and {Re Fiorentin}, P. and {Regibo}, S. and {Reyl{\'e}}, C. and {Ripepi}, V. and {Riva}, A. and {Rixon}, G. and {Robichon}, N. and {Robin}, C. and {Roelens}, M. and {Rohrbasser}, L. and {Romero-G{\'o}mez}, M. and {Rowell}, N. and {Royer}, F. and {Rybicki}, K.~A. and {Sadowski}, G. and {Sagrist{\`a} Sell{\'e}s}, A. and {Sahlmann}, J. and {Salgado}, J. and {Salguero}, E. and {Samaras}, N. and {Sanchez Gimenez}, V. and {Sanna}, N. and {Santove{\~n}a}, R. and {Sarasso}, M. and {Schultheis}, M. and {Sciacca}, E. and {Segol}, M. and {Segovia}, J.~C. and {S{\'e}gransan}, D. and {Semeux}, D. and {Shahaf}, S. and {Siddiqui}, H.~I. and {Siebert}, A. and {Siltala}, L. and {Slezak}, E. and {Smart}, R.~L. and {Solano}, E. and {Solitro}, F. and {Souami}, D. and {Souchay}, J. and {Spagna}, A. and {Spoto}, F. and {Steele}, I.~A. and {Steidelm{\"u}ller}, H. and {Stephenson}, C.~A. and {S{\"u}veges}, M. and {Szabados}, L. and {Szegedi-Elek}, E. and {Taris}, F. and {Tauran}, G. and {Taylor}, M.~B. and {Teixeira}, R. and {Thuillot}, W. and {Tonello}, N. and {Torra}, F. and {Torra}, J. and {Turon}, C. and {Unger}, N. and {Vaillant}, M. and {van Dillen}, E. and {Vanel}, O. and {Vecchiato}, A. and {Viala}, Y. and {Vicente}, D. and {Voutsinas}, S. and {Weiler}, M. and {Wevers}, T. and {Wyrzykowski}, {\L}. and {Yoldas}, A. and {Yvard}, P. and {Zhao}, H. and {Zorec}, J. and {Zucker}, S. and {Zurbach}, C. and {Zwitter}, T.},
        title = "{Gaia Early Data Release 3. Summary of the contents and survey properties}",
      journal = {\aap},
         year = 2021,
        month = may,
       volume = {649},
          eid = {A1},
        pages = {A1},
          doi = {10.1051/0004-6361/202039657},
archivePrefix = {arXiv},
       eprint = {2012.01533},
 primaryClass = {astro-ph.GA},
       adsurl = {https://ui.adsabs.harvard.edu/abs/2021A&A...649A...1G}
}

@ARTICLE{kurucz,
       author = {{Kurucz}, Robert L.},
        title = "{ATLAS12, SYNTHE, ATLAS9, WIDTH9, et cetera}",
      journal = {Memorie della Societa Astronomica Italiana Supplementi},
         year = 2005,
        month = jan,
       volume = {8},
        pages = {14},
       adsurl = {https://ui.adsabs.harvard.edu/abs/2005MSAIS...8...14K}
}

@ARTICLE{garro2026,
       author = {{Garro}, Elisa R. and {Massari}, Davide and {Fern{\'a}ndez-Trincado}, Jos{\'e} G. and {Ceccarelli}, Edoardo and {Sneden}, Chris and {Aguado-Agelet}, Fernando and {Af{\textcommabelow s}ar}, Melike and {Bellazzini}, Michele and {Guer{\c{c}}o}, Rafael and {Minniti}, Dante and et al.},
        title = "{HST+IGRINS synergy to characterise the newly discovered metal-rich bulge globular cluster Patchick 126}",
      journal = {\aap},
         year = 2026,
        month = feb,
       volume = {706},
          eid = {A338},
        pages = {A338},
          doi = {10.1051/0004-6361/202558157},
archivePrefix = {arXiv},
       eprint = {2601.03901},
 primaryClass = {astro-ph.GA},
       adsurl = {https://ui.adsabs.harvard.edu/abs/2026A&A...706A.338G}
}

@ARTICLE{martin2022,
       author = {{Martin}, Nicolas F. and {Venn}, Kim A. and {Aguado}, David S. and {Starkenburg}, Else and {Gonz{\'a}lez Hern{\'a}ndez}, Jonay I. and {Ibata}, Rodrigo A. and {Bonifacio}, Piercarlo and {Caffau}, Elisabetta and {Sestito}, Federico and {Arentsen}, Anke and et al.},
        title = "{A stellar stream remnant of a globular cluster below the metallicity floor}",
      journal = {\nat},
         year = 2022,
        month = jan,
       volume = {601},
       number = {7891},
        pages = {45-48},
          doi = {10.1038/s41586-021-04162-2},
archivePrefix = {arXiv},
       eprint = {2201.01309},
 primaryClass = {astro-ph.GA},
       adsurl = {https://ui.adsabs.harvard.edu/abs/2022Natur.601...45M}
}

@ARTICLE{gonzalez2020,
       author = {{Gonzalez}, O.~A. and {Mucciarelli}, A. and {Origlia}, L. and {Schultheis}, M. and {Caffau}, E. and {Di Matteo}, P. and {Randich}, S. and {Recio-Blanco}, A. and {Zoccali}, M. and {Bonifacio}, P. and et al.},
        title = "{MOONS Surveys of the Milky Way and its Satellites}",
      journal = {The Messenger},
         year = 2020,
        month = jun,
       volume = {180},
        pages = {18-23},
          doi = {10.18727/0722-6691/5196},
archivePrefix = {arXiv},
       eprint = {2009.00635},
 primaryClass = {astro-ph.GA},
       adsurl = {https://ui.adsabs.harvard.edu/abs/2020Msngr.180...18G}
}

@ARTICLE{hidalgo2018,
       author = {{Hidalgo}, Sebastian L. and {Pietrinferni}, Adriano and {Cassisi}, Santi and {Salaris}, Maurizio and {Mucciarelli}, Alessio and {Savino}, Alessandro and {Aparicio}, Antonio and {Silva Aguirre}, Victor and {Verma}, Kuldeep},
        title = "{The Updated BaSTI Stellar Evolution Models and Isochrones. I. Solar-scaled Calculations}",
      journal = {\apj},
         year = 2018,
        month = apr,
       volume = {856},
       number = {2},
          eid = {125},
        pages = {125},
          doi = {10.3847/1538-4357/aab158},
archivePrefix = {arXiv},
       eprint = {1802.07319},
 primaryClass = {astro-ph.GA},
       adsurl = {https://ui.adsabs.harvard.edu/abs/2018ApJ...856..125H}
}

@ARTICLE{vasiliev2021,
       author = {{Vasiliev}, Eugene and {Belokurov}, Vasily and {Erkal}, Denis},
        title = "{Tango for three: Sagittarius, LMC, and the Milky Way}",
      journal = {\mnras},
         year = 2021,
        month = feb,
       volume = {501},
       number = {2},
        pages = {2279-2304},
          doi = {10.1093/mnras/staa3673},
archivePrefix = {arXiv},
       eprint = {2009.10726},
 primaryClass = {astro-ph.GA},
       adsurl = {https://ui.adsabs.harvard.edu/abs/2021MNRAS.501.2279V}
}

@ARTICLE{pace2023,
       author = {{Pace}, Andrew B. and {Koposov}, Sergey E. and {Walker}, Matthew G. and {Caldwell}, Nelson and {Mateo}, Mario and {Olszewski}, Edward W. and {Roederer}, Ian U. and {Bailey}, John I. and {Belokurov}, Vasily and {Kuehn}, Kyler and et al.},
        title = "{The kinematics, metallicities, and orbits of six recently discovered Galactic star clusters with Magellan/M2FS spectroscopy}",
      journal = {\mnras},
         year = 2023,
        month = nov,
       volume = {526},
       number = {1},
        pages = {1075-1094},
          doi = {10.1093/mnras/stad2760},
archivePrefix = {arXiv},
       eprint = {2304.06904},
 primaryClass = {astro-ph.GA},
       adsurl = {https://ui.adsabs.harvard.edu/abs/2023MNRAS.526.1075P}
}

@ARTICLE{mucciarelli2026,
       author = {{Mucciarelli}, A. and {Bonifacio}, P. and {Lardo}, C.},
        title = "{KOALA, a new ATLAS9 database: I. Model atmospheres, opacities, fluxes, bolometric corrections, magnitudes, and colours}",
      journal = {\aap},
         year = 2026,
        month = jan,
       volume = {705},
          eid = {A134},
        pages = {A134},
          doi = {10.1051/0004-6361/202557640},
archivePrefix = {arXiv},
       eprint = {2511.10737},
 primaryClass = {astro-ph.SR},
       adsurl = {https://ui.adsabs.harvard.edu/abs/2026A&A...705A.134M}
}

@ARTICLE{vasiliev19,
       author = {{Vasiliev}, Eugene},
        title = "{AGAMA: action-based galaxy modelling architecture}",
      journal = {\mnras},
         year = 2019,
        month = jan,
       volume = {482},
       number = {2},
        pages = {1525-1544},
          doi = {10.1093/mnras/sty2672},
archivePrefix = {arXiv},
       eprint = {1802.08239},
 primaryClass = {astro-ph.GA},
       adsurl = {https://ui.adsabs.harvard.edu/abs/2019MNRAS.482.1525V}
}

@ARTICLE{paust2014,
       author = {{Paust}, Nathaniel and {Wilson}, Danielle and {van Belle}, Gerard},
        title = "{Reinvestigating the Clusters Koposov 1 and 2}",
      journal = {\aj},
         year = 2014,
        month = jul,
       volume = {148},
       number = {1},
          eid = {19},
        pages = {19},
          doi = {10.1088/0004-6256/148/1/19},
       adsurl = {https://ui.adsabs.harvard.edu/abs/2014AJ....148...19P}
}

@ARTICLE{koposov2007,
       author = {{Koposov}, S. and {de Jong}, J.~T.~A. and {Belokurov}, V. and {Rix}, H.-W. and {Zucker}, D.~B. and {Evans}, N.~W. and {Gilmore}, G. and {Irwin}, M.~J. and {Bell}, E.~F.},
        title = "{The Discovery of Two Extremely Low Luminosity Milky Way Globular Clusters}",
      journal = {\apj},
         year = 2007,
        month = nov,
       volume = {669},
       number = {1},
        pages = {337-342},
          doi = {10.1086/521422},
archivePrefix = {arXiv},
       eprint = {0706.0019},
 primaryClass = {astro-ph},
       adsurl = {https://ui.adsabs.harvard.edu/abs/2007ApJ...669..337K}
}

@ARTICLE{callingham2022,
       author = {{Callingham}, Thomas M. and {Cautun}, Marius and {Deason}, Alis J. and {Frenk}, Carlos S. and {Grand}, Robert J.~J. and {Marinacci}, Federico},
        title = "{The chemo-dynamical groups of Galactic globular clusters}",
      journal = {\mnras},
         year = 2022,
        month = jul,
       volume = {513},
       number = {3},
        pages = {4107-4129},
          doi = {10.1093/mnras/stac1145},
archivePrefix = {arXiv},
       eprint = {2202.00591},
 primaryClass = {astro-ph.GA},
       adsurl = {https://ui.adsabs.harvard.edu/abs/2022MNRAS.513.4107C}
}

@ARTICLE{vandenberg2013,
       author = {{VandenBerg}, Don A. and {Brogaard}, K. and {Leaman}, R. and {Casagrande}, L.},
        title = "{The Ages of 55 Globular Clusters as Determined Using an Improved $\Delta V^{HB}_{TO}$ Method along with Color-Magnitude Diagram Constraints, and Their Implications for Broader Issues}",
      journal = {\apj},
         year = 2013,
        month = oct,
       volume = {775},
       number = {2},
          eid = {134},
        pages = {134},
          doi = {10.1088/0004-637X/775/2/134},
archivePrefix = {arXiv},
       eprint = {1308.2257},
 primaryClass = {astro-ph.GA},
       adsurl = {https://ui.adsabs.harvard.edu/abs/2013ApJ...775..134V}
}

@ARTICLE{belokurov&kravstov2023,
       author = {{Belokurov}, Vasily and {Kravtsov}, Andrey},
        title = "{Nitrogen enrichment and clustered star formation at the dawn of the Galaxy}",
      journal = {\mnras},
         year = 2023,
        month = nov,
       volume = {525},
       number = {3},
        pages = {4456-4473},
          doi = {10.1093/mnras/stad2241},
archivePrefix = {arXiv},
       eprint = {2306.00060},
 primaryClass = {astro-ph.GA},
       adsurl = {https://ui.adsabs.harvard.edu/abs/2023MNRAS.525.4456B}
}

@ARTICLE{schiavon2017,
       author = {{Schiavon}, Ricardo P. and {Zamora}, Olga and {Carrera}, Ricardo and {Lucatello}, Sara and {Robin}, A.~C. and {Ness}, Melissa and {Martell}, Sarah L. and {Smith}, Verne V. and {Garc{\'\i}a-Hern{\'a}ndez}, D.~A. and {Manchado}, Arturo and {Sch{\"o}nrich}, Ralph and {Bastian}, Nate and {Chiappini}, Cristina and {Shetrone}, Matthew and {Mackereth}, J. Ted and {Williams}, Rob A. and {M{\'e}sz{\'a}ros}, Szabolcs and {Allende Prieto}, Carlos and {Anders}, Friedrich and {Bizyaev}, Dmitry and {Beers}, Timothy C. and {Chojnowski}, S. Drew and {Cunha}, Katia and {Epstein}, Courtney and {Frinchaboy}, Peter M. and {Garc{\'\i}a P{\'e}rez}, Ana E. and {Hearty}, Fred R. and {Holtzman}, Jon A. and {Johnson}, Jennifer A. and {Kinemuchi}, Karen and {Majewski}, Steven R. and {Muna}, Demitri and {Nidever}, David L. and {Nguyen}, Duy Cuong and {O'Connell}, Robert W. and {Oravetz}, Daniel and {Pan}, Kaike and {Pinsonneault}, Marc and {Schneider}, Donald P. and {Schultheis}, Matthias and {Simmons}, Audrey and {Skrutskie}, Michael F. and {Sobeck}, Jennifer and {Wilson}, John C. and {Zasowski}, Gail},
        title = "{Chemical tagging with APOGEE: discovery of a large population of N-rich stars in the inner Galaxy}",
      journal = {\mnras},
         year = 2017,
        month = feb,
       volume = {465},
       number = {1},
        pages = {501-524},
          doi = {10.1093/mnras/stw2162},
archivePrefix = {arXiv},
       eprint = {1606.05651},
 primaryClass = {astro-ph.GA},
       adsurl = {https://ui.adsabs.harvard.edu/abs/2017MNRAS.465..501S}
}

@ARTICLE{monty2024,
       author = {{Monty}, Stephanie and {Belokurov}, Vasily and {Sanders}, Jason L. and {Hansen}, Terese T. and {Sakari}, Charli M. and {McKenzie}, Madeleine and {Myeong}, GyuChul and {Davies}, Elliot Y. and {Ardern-Arentsen}, Anke and {Massari}, Davide},
        title = "{The ratio of [Eu/{\ensuremath{\alpha}}] differentiates accreted/in situ Milky Way stars across metallicities, as indicated by both field stars and globular clusters}",
      journal = {\mnras},
         year = 2024,
        month = sep,
       volume = {533},
       number = {2},
        pages = {2420-2440},
          doi = {10.1093/mnras/stae1895},
archivePrefix = {arXiv},
       eprint = {2405.08963},
 primaryClass = {astro-ph.GA},
       adsurl = {https://ui.adsabs.harvard.edu/abs/2024MNRAS.533.2420M}
}

@ARTICLE{massari2025,
       author = {{Massari}, D. and {Bellazzini}, M. and {Libralato}, M. and {Bellini}, A. and {Dalessandro}, E. and {Ceccarelli}, E. and {Aguado-Agelet}, F. and {Cassisi}, S. and {Gallart}, C. and {Monelli}, M. and {Mucciarelli}, A. and {Pancino}, E. and {Salaris}, M. and {Saracino}, S. and {Dodd}, E. and {Ferraro}, F.~R. and {Garro}, E.~R. and {Lanzoni}, B. and {Pascale}, R. and {Rosignoli}, L.},
        title = "{The Hubble Missing Globular Cluster Survey: I. Survey overview and the first precise age estimate for ESO452-11 and 2MASS-GC01}",
      journal = {\aap},
         year = 2025,
        month = jun,
       volume = {698},
          eid = {A197},
        pages = {A197},
          doi = {10.1051/0004-6361/202554007},
archivePrefix = {arXiv},
       eprint = {2502.01741},
 primaryClass = {astro-ph.GA},
       adsurl = {https://ui.adsabs.harvard.edu/abs/2025A&A...698A.197M}
}

@ARTICLE{adamo2024,
       author = {{Adamo}, Angela and {Bradley}, Larry D. and {Vanzella}, Eros and {Claeyssens}, Ad{\'e}la{\"\i}de and {Welch}, Brian and {Diego}, Jose M. and {Mahler}, Guillaume and {Oguri}, Masamune and {Sharon}, Keren and {Abdurro'uf} and {Hsiao}, Tiger Yu-Yang and {Xu}, Xinfeng and {Messa}, Matteo and {Lassen}, Augusto E. and {Zackrisson}, Erik and {Brammer}, Gabriel and {Coe}, Dan and {Kokorev}, Vasily and {Ricotti}, Massimo and {Zitrin}, Adi and {Fujimoto}, Seiji and {Inoue}, Akio K. and {Resseguier}, Tom and {Rigby}, Jane R. and {Jim{\'e}nez-Teja}, Yolanda and {Windhorst}, Rogier A. and {Hashimoto}, Takuya and {Tamura}, Yoichi},
        title = "{Bound star clusters observed in a lensed galaxy 460 Myr after the Big Bang}",
      journal = {\nat},
         year = 2024,
        month = aug,
       volume = {632},
       number = {8025},
        pages = {513-516},
          doi = {10.1038/s41586-024-07703-7},
archivePrefix = {arXiv},
       eprint = {2401.03224},
 primaryClass = {astro-ph.GA},
       adsurl = {https://ui.adsabs.harvard.edu/abs/2024Natur.632..513A}
}

@ARTICLE{forbes&bridges2010,
       author = {{Forbes}, Duncan A. and {Bridges}, Terry},
        title = "{Accreted versus in situ Milky Way globular clusters}",
      journal = {\mnras},
         year = 2010,
        month = may,
       volume = {404},
       number = {3},
        pages = {1203-1214},
          doi = {10.1111/j.1365-2966.2010.16373.x},
archivePrefix = {arXiv},
       eprint = {1001.4289},
 primaryClass = {astro-ph.GA},
       adsurl = {https://ui.adsabs.harvard.edu/abs/2010MNRAS.404.1203F}
}

@ARTICLE{horta2020,
       author = {{Horta}, Danny and {Schiavon}, Ricardo P. and {Mackereth}, J. Ted and {Beers}, Timothy C. and {Fern{\'a}ndez-Trincado}, Jos{\'e} G. and {Frinchaboy}, Peter M. and {Garc{\'\i}a-Hern{\'a}ndez}, D.~A. and {Geisler}, Doug and {Hasselquist}, Sten and {J{\"o}nsson}, Henrik and {Lane}, Richard R. and {Majewski}, Steven R. and {M{\'e}sz{\'a}ros}, Szabolcs and {Bidin}, Christian Moni and {Nataf}, David M. and {Roman-Lopes}, Alexandre and {Nitschelm}, Christian and {Vargas-Gonz{\'a}lez}, J. and {Zasowski}, Gail},
        title = "{The chemical compositions of accreted and in situ galactic globular clusters according to SDSS/APOGEE}",
      journal = {\mnras},
         year = 2020,
        month = apr,
       volume = {493},
       number = {3},
        pages = {3363-3378},
          doi = {10.1093/mnras/staa478},
archivePrefix = {arXiv},
       eprint = {2001.03177},
 primaryClass = {astro-ph.GA},
       adsurl = {https://ui.adsabs.harvard.edu/abs/2020MNRAS.493.3363H}
}

@ARTICLE{searle&zinn1978,
       author = {{Searle}, L. and {Zinn}, R.},
        title = "{Composition of halo clusters and the formation of the galactic halo.}",
      journal = {\apj},
         year = 1978,
        month = oct,
       volume = {225},
        pages = {357-379},
          doi = {10.1086/156499},
       adsurl = {https://ui.adsabs.harvard.edu/abs/1978ApJ...225..357S}
}

@ARTICLE{vasiliev&baumgardt2021,
       author = {{Vasiliev}, Eugene and {Baumgardt}, Holger},
        title = "{Gaia EDR3 view on galactic globular clusters}",
      journal = {\mnras},
         year = 2021,
        month = aug,
       volume = {505},
       number = {4},
        pages = {5978-6002},
          doi = {10.1093/mnras/stab1475},
archivePrefix = {arXiv},
       eprint = {2102.09568},
 primaryClass = {astro-ph.GA},
       adsurl = {https://ui.adsabs.harvard.edu/abs/2021MNRAS.505.5978V}
}

@INPROCEEDINGS{pogge2010,
       author = {{Pogge}, R.~W. and {Atwood}, B. and {Brewer}, D.~F. and {Byard}, P.~L. and {Derwent}, M.~A. and {Gonzalez}, R. and {Martini}, P. and {Mason}, J.~A. and {O'Brien}, T.~P. and {Osmer}, P.~S. and {Pappalardo}, D.~P. and {Steinbrecher}, D.~P. and {Teiga}, E.~J. and {Zhelem}, R.},
        title = "{The multi-object double spectrographs for the Large Binocular Telescope}",
    booktitle = {Ground-based and Airborne Instrumentation for Astronomy III},
         year = 2010,
       editor = {{McLean}, Ian S. and {Ramsay}, Suzanne K. and {Takami}, Hideki},
       series = {Society of Photo-Optical Instrumentation Engineers (SPIE) Conference Series},
       volume = {7735},
        month = jul,
          eid = {77350A},
        pages = {77350A},
          doi = {10.1117/12.857215},
       adsurl = {https://ui.adsabs.harvard.edu/abs/2010SPIE.7735E..0AP}
}

@ARTICLE{hanuschik2003,
       author = {{Hanuschik}, R.~W.},
        title = "{A flux-calibrated, high-resolution atlas of optical sky emission from UVES}",
      journal = {\aap},
         year = 2003,
        month = sep,
       volume = {407},
        pages = {1157-1164},
          doi = {10.1051/0004-6361:20030885},
       adsurl = {https://ui.adsabs.harvard.edu/abs/2003A&A...407.1157H}
}

@ARTICLE{edlen1966,
       author = {{Edl{\'e}n}, Bengt},
        title = "{The Refractive Index of Air}",
      journal = {Metrologia},
         year = 1966,
        month = apr,
       volume = {2},
       number = {2},
        pages = {71-80},
          doi = {10.1088/0026-1394/2/2/002},
       adsurl = {https://ui.adsabs.harvard.edu/abs/1966Metro...2...71E}
}

@ARTICLE{gargiulo22,
       author = {{Gargiulo}, A. and {Fumana}, M. and {Bisogni}, S. and {Franzetti}, P. and {Cassar{\`a}}, L.~P. and {Garilli}, B. and {Scodeggio}, M. and {Vietri}, G.},
        title = "{SIPGI: an interactive pipeline for spectroscopic data reduction}",
      journal = {\mnras},
         year = 2022,
        month = aug,
       volume = {514},
       number = {2},
        pages = {2902-2914},
          doi = {10.1093/mnras/stac1065},
archivePrefix = {arXiv},
       eprint = {2209.05441},
 primaryClass = {astro-ph.IM},
       adsurl = {https://ui.adsabs.harvard.edu/abs/2022MNRAS.514.2902G}
}

@ARTICLE{mcmillan17,
       author = {{McMillan}, Paul J.},
        title = "{The mass distribution and gravitational potential of the Milky Way}",
      journal = {\mnras},
         year = 2017,
        month = feb,
       volume = {465},
       number = {1},
        pages = {76-94},
          doi = {10.1093/mnras/stw2759},
archivePrefix = {arXiv},
       eprint = {1608.00971},
 primaryClass = {astro-ph.GA},
       adsurl = {https://ui.adsabs.harvard.edu/abs/2017MNRAS.465...76M}
}

@ARTICLE{mucciarelli21,
       author = {{Mucciarelli}, A. and {Bellazzini}, M. and {Massari}, D.},
        title = "{Exploiting the Gaia EDR3 photometry to derive stellar temperatures}",
      journal = {\aap},
         year = 2021,
        month = sep,
       volume = {653},
          eid = {A90},
        pages = {A90},
          doi = {10.1051/0004-6361/202140979},
archivePrefix = {arXiv},
       eprint = {2106.03882},
 primaryClass = {astro-ph.SR},
       adsurl = {https://ui.adsabs.harvard.edu/abs/2021A&A...653A..90M}
}

@ARTICLE{gravitycollaboration18,
       author = {{GRAVITY Collaboration} and {Abuter}, R. and {Amorim}, A. and {Anugu}, N. and {Baub{\"o}ck}, M. and {Benisty}, M. and {Berger}, J.~P. and {Blind}, N. and {Bonnet}, H. and {Brandner}, W. and {Buron}, A. and {Collin}, C. and {Chapron}, F. and {Cl{\'e}net}, Y. and {Coud{\'e} Du Foresto}, V. and {de Zeeuw}, P.~T. and {Deen}, C. and {Delplancke-Str{\"o}bele}, F. and {Dembet}, R. and {Dexter}, J. and {Duvert}, G. and {Eckart}, A. and {Eisenhauer}, F. and {Finger}, G. and {F{\"o}rster Schreiber}, N.~M. and {F{\'e}dou}, P. and {Garcia}, P. and {Garcia Lopez}, R. and {Gao}, F. and {Gendron}, E. and {Genzel}, R. and {Gillessen}, S. and {Gordo}, P. and {Habibi}, M. and {Haubois}, X. and {Haug}, M. and {Hau{\ss}mann}, F. and {Henning}, Th. and {Hippler}, S. and {Horrobin}, M. and {Hubert}, Z. and {Hubin}, N. and {Jimenez Rosales}, A. and {Jochum}, L. and {Jocou}, K. and {Kaufer}, A. and {Kellner}, S. and {Kendrew}, S. and {Kervella}, P. and {Kok}, Y. and {Kulas}, M. and {Lacour}, S. and {Lapeyr{\`e}re}, V. and {Lazareff}, B. and {Le Bouquin}, J. -B. and {L{\'e}na}, P. and {Lippa}, M. and {Lenzen}, R. and {M{\'e}rand}, A. and {M{\"u}ler}, E. and {Neumann}, U. and {Ott}, T. and {Palanca}, L. and {Paumard}, T. and {Pasquini}, L. and {Perraut}, K. and {Perrin}, G. and {Pfuhl}, O. and {Plewa}, P.~M. and {Rabien}, S. and {Ram{\'\i}rez}, A. and {Ramos}, J. and {Rau}, C. and {Rodr{\'\i}guez-Coira}, G. and {Rohloff}, R. -R. and {Rousset}, G. and {Sanchez-Bermudez}, J. and {Scheithauer}, S. and {Sch{\"o}ller}, M. and {Schuler}, N. and {Spyromilio}, J. and {Straub}, O. and {Straubmeier}, C. and {Sturm}, E. and {Tacconi}, L.~J. and {Tristram}, K.~R.~W. and {Vincent}, F. and {von Fellenberg}, S. and {Wank}, I. and {Waisberg}, I. and {Widmann}, F. and {Wieprecht}, E. and {Wiest}, M. and {Wiezorrek}, E. and {Woillez}, J. and {Yazici}, S. and {Ziegler}, D. and {Zins}, G.},
        title = "{Detection of the gravitational redshift in the orbit of the star S2 near the Galactic centre massive black hole}",
      journal = {\aap},
         year = 2018,
        month = jul,
       volume = {615},
          eid = {L15},
        pages = {L15},
          doi = {10.1051/0004-6361/201833718},
archivePrefix = {arXiv},
       eprint = {1807.09409},
 primaryClass = {astro-ph.GA},
       adsurl = {https://ui.adsabs.harvard.edu/abs/2018A&A...615L..15G}
}

@ARTICLE{bennetandbovy19,
       author = {{Bennett}, Morgan and {Bovy}, Jo},
        title = "{Vertical waves in the solar neighbourhood in Gaia DR2}",
      journal = {\mnras},
         year = 2019,
        month = jan,
       volume = {482},
       number = {1},
        pages = {1417-1425},
          doi = {10.1093/mnras/sty2813},
archivePrefix = {arXiv},
       eprint = {1809.03507},
 primaryClass = {astro-ph.GA},
       adsurl = {https://ui.adsabs.harvard.edu/abs/2019MNRAS.482.1417B}
}

@ARTICLE{drimmelandpoggio18,
       author = {{Drimmel}, Ronald and {Poggio}, Eloisa},
        title = "{On the Solar Velocity}",
      journal = {Research Notes of the American Astronomical Society},
         year = 2018,
        month = nov,
       volume = {2},
       number = {4},
          eid = {210},
        pages = {210},
          doi = {10.3847/2515-5172/aaef8b},
       adsurl = {https://ui.adsabs.harvard.edu/abs/2018RNAAS...2..210D}
}

@ARTICLE{ibata94,
       author = {{Ibata}, R.~A. and {Gilmore}, G. and {Irwin}, M.~J.},
        title = "{A dwarf satellite galaxy in Sagittarius}",
      journal = {\nat},
         year = 1994,
        month = jul,
       volume = {370},
       number = {6486},
        pages = {194-196},
          doi = {10.1038/370194a0},
       adsurl = {https://ui.adsabs.harvard.edu/abs/1994Natur.370..194I}
}

@ARTICLE{helmi2018,
       author = {{Helmi}, Amina and {Babusiaux}, Carine and {Koppelman}, Helmer H. and {Massari}, Davide and {Veljanoski}, Jovan and {Brown}, Anthony G.~A.},
        title = "{The merger that led to the formation of the Milky Way's inner stellar halo and thick disk}",
      journal = {\nat},
         year = 2018,
        month = oct,
       volume = {563},
       number = {7729},
        pages = {85-88},
          doi = {10.1038/s41586-018-0625-x},
archivePrefix = {arXiv},
       eprint = {1806.06038},
 primaryClass = {astro-ph.GA},
       adsurl = {https://ui.adsabs.harvard.edu/abs/2018Natur.563...85H}
}

@ARTICLE{belokurov2018,
       author = {{Belokurov}, V. and {Erkal}, D. and {Evans}, N.~W. and {Koposov}, S.~E. and {Deason}, A.~J.},
        title = "{Co-formation of the disc and the stellar halo}",
      journal = {\mnras},
         year = 2018,
        month = jul,
       volume = {478},
       number = {1},
        pages = {611-619},
          doi = {10.1093/mnras/sty982},
archivePrefix = {arXiv},
       eprint = {1802.03414},
 primaryClass = {astro-ph.GA},
       adsurl = {https://ui.adsabs.harvard.edu/abs/2018MNRAS.478..611B}
}

@ARTICLE{kruijssen2020,
       author = {{Kruijssen}, J.~M. Diederik and {Pfeffer}, Joel L. and {Chevance}, M{\'e}lanie and {Bonaca}, Ana and {Trujillo-Gomez}, Sebastian and {Bastian}, Nate and {Reina-Campos}, Marta and {Crain}, Robert A. and {Hughes}, Meghan E.},
        title = "{Kraken reveals itself - the merger history of the Milky Way reconstructed with the E-MOSAICS simulations}",
      journal = {\mnras},
         year = 2020,
        month = oct,
       volume = {498},
       number = {2},
        pages = {2472-2491},
          doi = {10.1093/mnras/staa2452},
archivePrefix = {arXiv},
       eprint = {2003.01119},
 primaryClass = {astro-ph.GA},
       adsurl = {https://ui.adsabs.harvard.edu/abs/2020MNRAS.498.2472K}
}

@ARTICLE{tonry&davies1979,
       author = {{Tonry}, J. and {Davis}, M.},
        title = "{A survey of galaxy redshifts. I. Data reduction techniques.}",
      journal = {\aj},
         year = 1979,
        month = oct,
       volume = {84},
        pages = {1511-1525},
          doi = {10.1086/112569},
       adsurl = {https://ui.adsabs.harvard.edu/abs/1979AJ.....84.1511T}
}

@ARTICLE{massari19,
       author = {{Massari}, D. and {Koppelman}, H.~H. and {Helmi}, A.},
        title = "{Origin of the system of globular clusters in the Milky Way}",
      journal = {\aap},
         year = 2019,
        month = oct,
       volume = {630},
          eid = {L4},
        pages = {L4},
          doi = {10.1051/0004-6361/201936135},
archivePrefix = {arXiv},
       eprint = {1906.08271},
 primaryClass = {astro-ph.GA},
       adsurl = {https://ui.adsabs.harvard.edu/abs/2019A&A...630L...4M}
}

@ARTICLE{riello2021,
       author = {{Riello}, M. and {De Angeli}, F. and {Evans}, D.~W. and {Montegriffo}, P. and {Carrasco}, J.~M. and {Busso}, G. and {Palaversa}, L. and {Burgess}, P.~W. and {Diener}, C. and {Davidson}, M. and {Rowell}, N. and {Fabricius}, C. and {Jordi}, C. and {Bellazzini}, M. and {Pancino}, E. and {Harrison}, D.~L. and {Cacciari}, C. and {van Leeuwen}, F. and {Hambly}, N.~C. and {Hodgkin}, S.~T. and {Osborne}, P.~J. and {Altavilla}, G. and {Barstow}, M.~A. and {Brown}, A.~G.~A. and {Castellani}, M. and {Cowell}, S. and {De Luise}, F. and {Gilmore}, G. and {Giuffrida}, G. and {Hidalgo}, S. and {Holland}, G. and {Marinoni}, S. and {Pagani}, C. and {Piersimoni}, A.~M. and {Pulone}, L. and {Ragaini}, S. and {Rainer}, M. and {Richards}, P.~J. and {Sanna}, N. and {Walton}, N.~A. and {Weiler}, M. and {Yoldas}, A.},
        title = "{Gaia Early Data Release 3. Photometric content and validation}",
      journal = {\aap},
         year = 2021,
        month = may,
       volume = {649},
          eid = {A3},
        pages = {A3},
          doi = {10.1051/0004-6361/202039587},
archivePrefix = {arXiv},
       eprint = {2012.01916},
 primaryClass = {astro-ph.IM},
       adsurl = {https://ui.adsabs.harvard.edu/abs/2021A&A...649A...3R}
}

@ARTICLE{horta2021,
       author = {{Horta}, Danny and {Schiavon}, Ricardo P. and {Mackereth}, J. Ted and {Pfeffer}, Joel and {Mason}, Andrew C. and {Kisku}, Shobhit and {Fragkoudi}, Francesca and {Allende Prieto}, Carlos and {Cunha}, Katia and {Hasselquist}, Sten and {Holtzman}, Jon and {Majewski}, Steven R. and {Nataf}, David and {O'Connell}, Robert W. and {Schultheis}, Mathias and {Smith}, Verne V.},
        title = "{Evidence from APOGEE for the presence of a major building block of the halo buried in the inner Galaxy}",
      journal = {\mnras},
         year = 2021,
        month = jan,
       volume = {500},
       number = {1},
        pages = {1385-1403},
          doi = {10.1093/mnras/staa2987},
archivePrefix = {arXiv},
       eprint = {2007.10374},
 primaryClass = {astro-ph.GA},
       adsurl = {https://ui.adsabs.harvard.edu/abs/2021MNRAS.500.1385H}
}

@ARTICLE{pietrinferni2021,
       author = {{Pietrinferni}, Adriano and {Hidalgo}, Sebastian and {Cassisi}, Santi and {Salaris}, Maurizio and {Savino}, Alessandro and {Mucciarelli}, Alessio and {Verma}, Kuldeep and {Silva Aguirre}, Victor and {Aparicio}, Antonio and {Ferguson}, Jason W.},
        title = "{Updated BaSTI Stellar Evolution Models and Isochrones. II. {\ensuremath{\alpha}}-enhanced Calculations}",
      journal = {\apj},
         year = 2021,
        month = feb,
       volume = {908},
       number = {1},
          eid = {102},
        pages = {102},
          doi = {10.3847/1538-4357/abd4d5},
archivePrefix = {arXiv},
       eprint = {2012.10085},
 primaryClass = {astro-ph.SR},
       adsurl = {https://ui.adsabs.harvard.edu/abs/2021ApJ...908..102P}
}

@ARTICLE{vanzella2017,
       author = {{Vanzella}, E. and {Calura}, F. and {Meneghetti}, M. and {Mercurio}, A. and {Castellano}, M. and {Caminha}, G.~B. and {Balestra}, I. and {Rosati}, P. and {Tozzi}, P. and {De Barros}, S. and {Grazian}, A. and {D'Ercole}, A. and {Ciotti}, L. and {Caputi}, K. and {Grillo}, C. and {Merlin}, E. and {Pentericci}, L. and {Fontana}, A. and {Cristiani}, S. and {Coe}, D.},
        title = "{Paving the way for the JWST: witnessing globular cluster formation at z > 3}",
      journal = {\mnras},
         year = 2017,
        month = jun,
       volume = {467},
       number = {4},
        pages = {4304-4321},
          doi = {10.1093/mnras/stx351},
archivePrefix = {arXiv},
       eprint = {1612.01526},
 primaryClass = {astro-ph.GA},
       adsurl = {https://ui.adsabs.harvard.edu/abs/2017MNRAS.467.4304V}
}

@ARTICLE{claeyssens2023,
       author = {{Claeyssens}, Ad{\'e}la{\"\i}de and {Adamo}, Angela and {Richard}, Johan and {Mahler}, Guillaume and {Messa}, Matteo and {Dessauges-Zavadsky}, Miroslava},
        title = "{Star formation at the smallest scales: a JWST study of the clump populations in SMACS0723}",
      journal = {\mnras},
         year = 2023,
        month = apr,
       volume = {520},
       number = {2},
        pages = {2180-2203},
          doi = {10.1093/mnras/stac3791},
archivePrefix = {arXiv},
       eprint = {2208.10450},
 primaryClass = {astro-ph.GA},
       adsurl = {https://ui.adsabs.harvard.edu/abs/2023MNRAS.520.2180C}
}

@ARTICLE{deleo2026_bo,
       author = {{De Leo}, M. and {Massari}, D. and {Bellazzini}, M. and {Mucciarelli}, A. and {Acosta-Tripailao}, B. and {Nipoti}, C.},
        title = "{Dynamical mirages: How bar-induced resonant trapping can mimic substructure clustering in dynamical parameter spaces}",
      journal = {\aap},
         year = 2026,
        month = mar,
       volume = {707},
          eid = {A310},
        pages = {A310},
          doi = {10.1051/0004-6361/202556723},
archivePrefix = {arXiv},
       eprint = {2511.05655},
 primaryClass = {astro-ph.GA},
       adsurl = {https://ui.adsabs.harvard.edu/abs/2026A&A...707A.310D}
}

%---------------------------------------------------
\begin{appendix}
\section{Targets information, additional CMDs and spectra}\label{app:A}
%-------------------------------------------------------------
%-------------------------- Table -------------------------
\begin{table*}
\caption{Information on the observations for target stars.}        
\label{tab:obs_stars}      
\centering          
\begin{tabular}{lcccccccc}  
\hline 
\hline      
Cluster & \textit{Gaia} DR3 & R.A. & Dec. & $G$ & $BP$ & $RP$ & S/N & S/N \\ 
 & \texttt{source\_id} & (deg) & (deg) & (mag)  & (mag) & (mag) & [$6500$ \r{A}] & [$8500$ \r{A})]    \\ 
\hline 
\multirow{4}{*}{Koposov 1} & 3919839462184529664 & 179.8069 & 12.2667 & 17.66 & 18.20 & 17.00 & 61 & 63 \\
  & 3919839767126601728 & 179.8297 & 12.2656 & 20.23 & 20.77 & 19.70 & 11 &  7 \\
  & 3919839767126647680 & 179.8255 & 12.2576 & 20.26 & 20.51 & 20.08 & 10 &  5 \\
  & 3919839771422278016 & 179.8270 & 12.2626 & 20.52 & 20.55 & 19.96 &  9 &  4 \\
\hline   
\multirow{4}{*}{Koposov 2} & 874275674394169984 & 119.5718 & 26.2599 & 18.61 & 19.03 & 18.05 & 83 & 68 \\
  & 874275609971796992 & 119.5770 & 26.2570 & 20.55 & 20.82 & 20.23 & 21 &  9 \\
  & 874269730159393536 & 119.5481 & 26.2379 & 20.45 & 20.92 & 19.98 & 15 &  8 \\
  & 874275850488194176 & 119.5530 & 26.2642 & 20.79 & 21.11 & 20.02 & 15 & 10 \\
\hline 
\multirow{4}{*}{Mu\~noz 1} & 1693370742839913088 & 225.4229 & 66.9687 & 19.06 & 19.12 & 19.14 & 33 & 18 \\
  & 1693394146117710592 & 225.4497 & 66.9692 & 19.22 & 19.29 & 19.21 & 30 & 14 \\
  & 1693394176181484672 & 225.4762 & 66.9767 & 20.30 & 20.96 & 19.87 & 14 & 10 \\
  & 1693393420267235840 & 225.4770 & 66.9673 & 20.73 & 21.14 & 20.41 &  9 &  4 \\
\hline 
\multirow{7}{*}{Pfleiderer 2} & 4172573170590468992 & 269.6989 & -5.0826 & 16.61 & 18.36 & 15.37 & 46 &  77 \\
  & 4175575730690574208 & 269.6486 & -5.0720 & 17.39 & 18.64 & 16.30 & 50 &  82 \\
  & 4172549561149372160 & 269.6726 & -5.1066 & 16.99 & 18.50 & 15.82 & 59 &  86 \\
  & 4172549836027324800 & 269.6482 & -5.0854 & 17.63 & 18.95 & 16.52 & 66 &  90 \\
  & 4172549840328133120 & 269.6419 & -5.0864 & 17.66 & 18.97 & 16.54 & 58 &  81 \\
  & 4175575863825402880 & 269.6781 & -5.0543 & 17.79 & 19.27 & 16.60 & 44 &  63 \\
  & 4172573269368904448 & 269.6809 & -5.0790 & 17.67 & 19.06 & 16.54 & 32 &  44 \\
\hline  
\multirow{4}{*}{RLGC2} & 4253618683784925568 & 281.3664 & -5.1897 & 17.39 & 20.21 & 15.87 & 62 & 120 \\
  & 4253618683784927872 & 281.3694 & -5.1931 & 17.92 & 20.74 & 16.40 & 45 & 112 \\
  & 4253618683784923392 & 281.3602 & -5.1919 & 18.28 & 20.51 & 16.54 & 33 & 118 \\
  & 4253618683784929536 & 281.3732 & -5.1947 & 18.15 & 20.92 & 16.70 & 44 & 122 \\
\hline 
\end{tabular}
\tablefoot{Cluster, \texttt{source\_id}, position, and magnitudes from \textit{Gaia} DR3, alongside S/N around the H$\alpha$ and CaT features. 
}
\end{table*}
%-------------------------------------------------------------
\begin{figure*}[!h]
\centering
\includegraphics[width=0.95\textwidth]{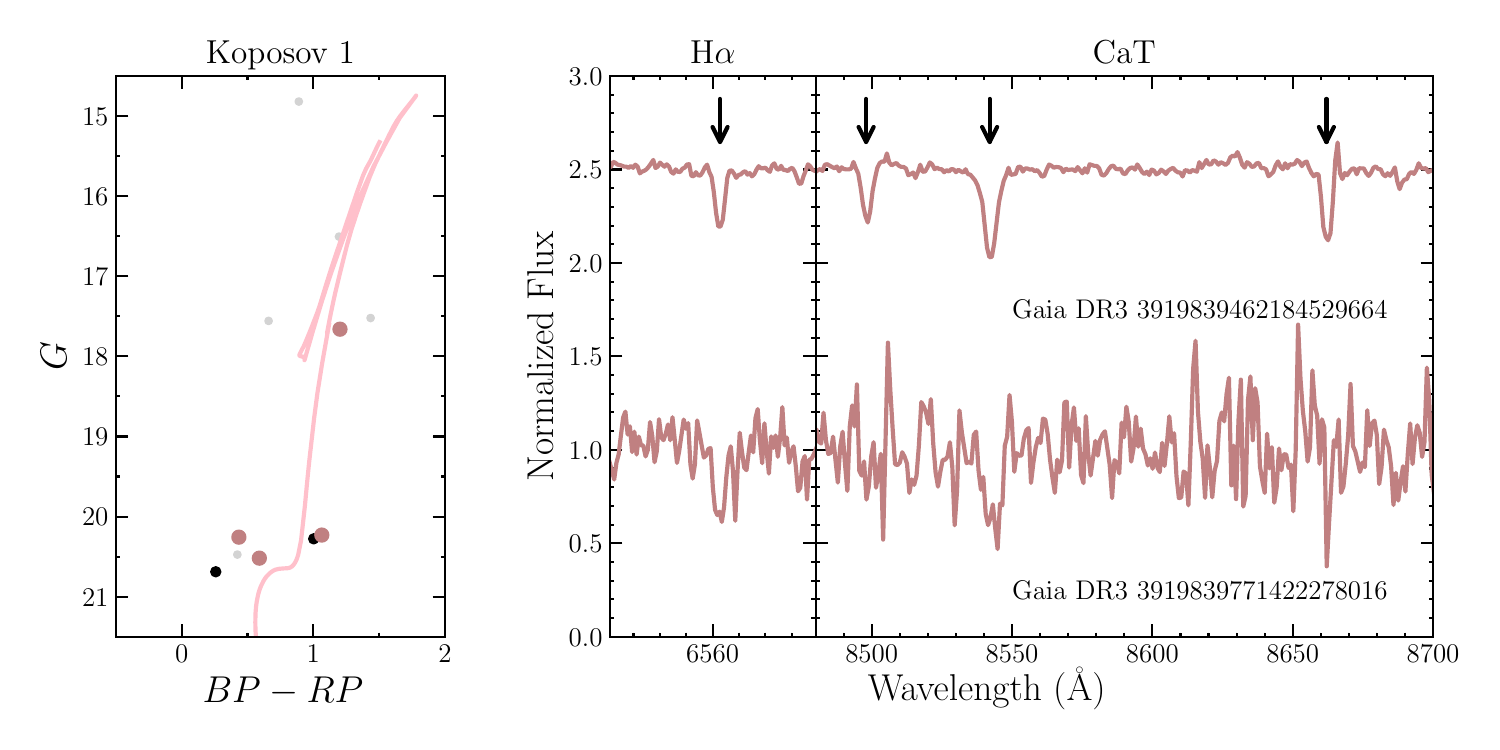} 
\includegraphics[width=0.95\textwidth]{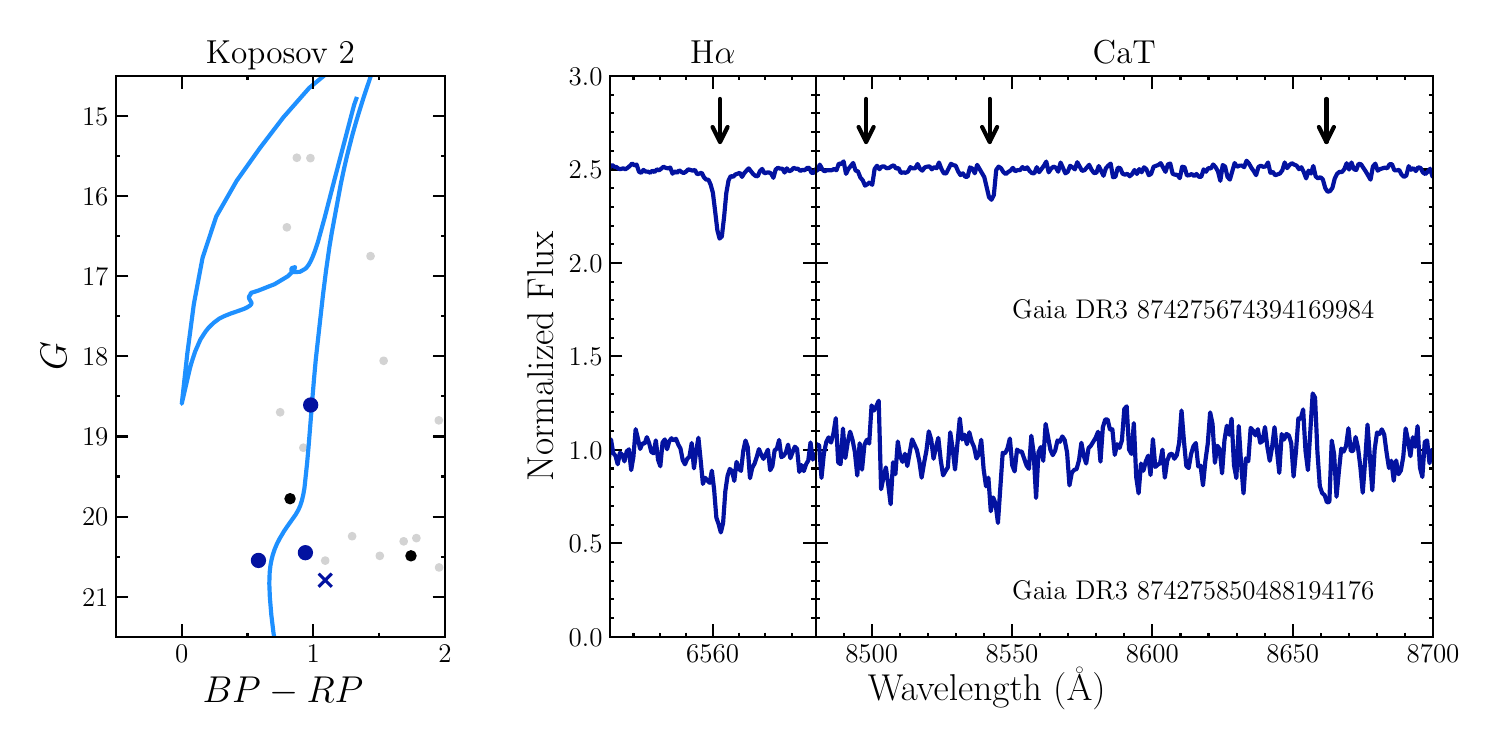}
\caption{Same as Fig. \ref{fig:data_Pfleiderer2}, but for Koposov 1 (pink) and Koposov 2 (blue). Reference BaSTI isochrone are 7 and 13.5 Gyr old, $\feh = -1.2$ dex and $\feh = -2.9$ dex, respectively, and $\alpha$-enanched ($\afeh = +0.4$ dex). We plot them assuming reddening from \citet{paust2014} for both clusters, and distances from \citet{paust2014} and \citet{cerny26} for Koposov 1 and Koposov 2. Stars in our sample identified as outliers in $\vlos$ are marked with a coloured cross, if present.}
\label{fig:data_2}%
\end{figure*}
%-------------------------------------------------------------
%-------------------------------------------------------------
\begin{figure*}[!h]
\centering
\includegraphics[width=0.95\textwidth]{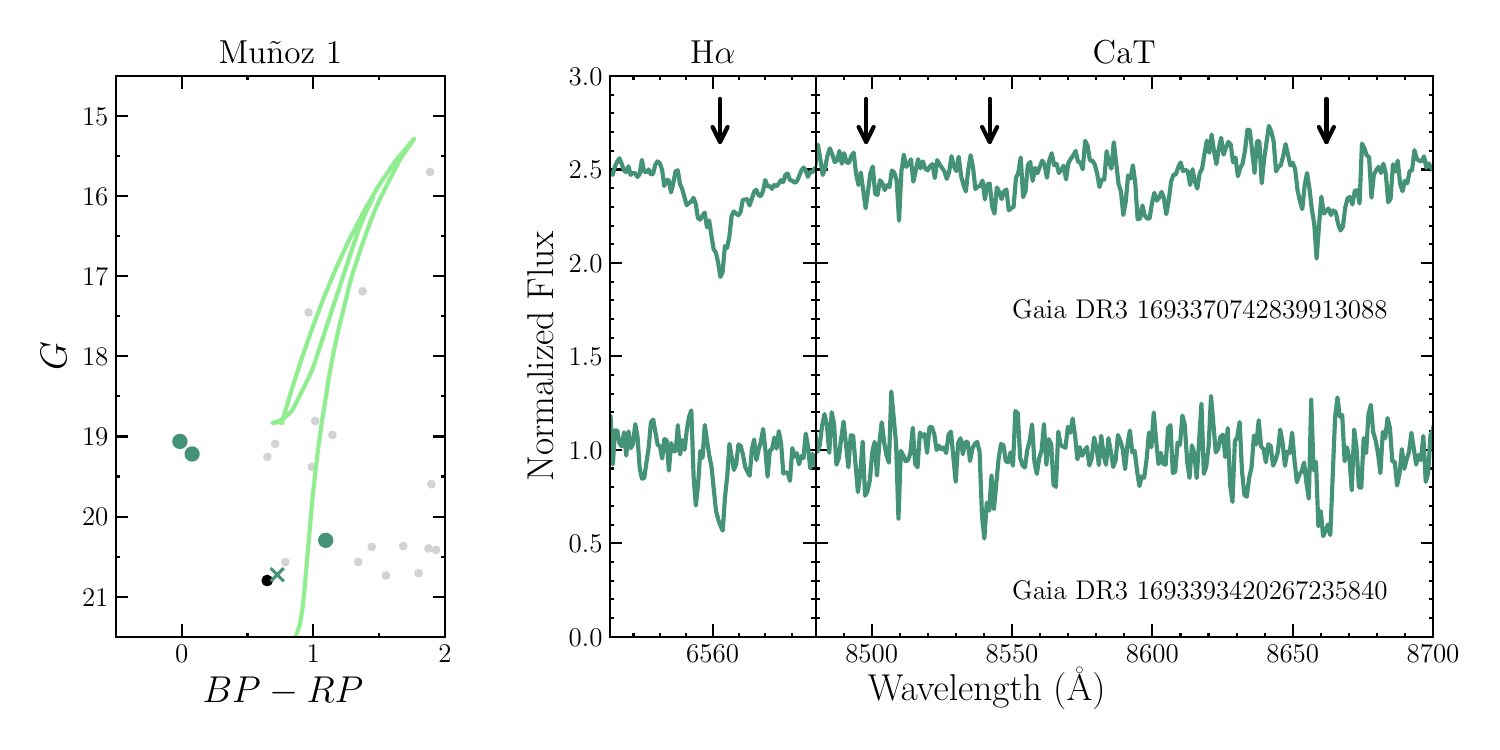} 
\includegraphics[width=0.95\textwidth]{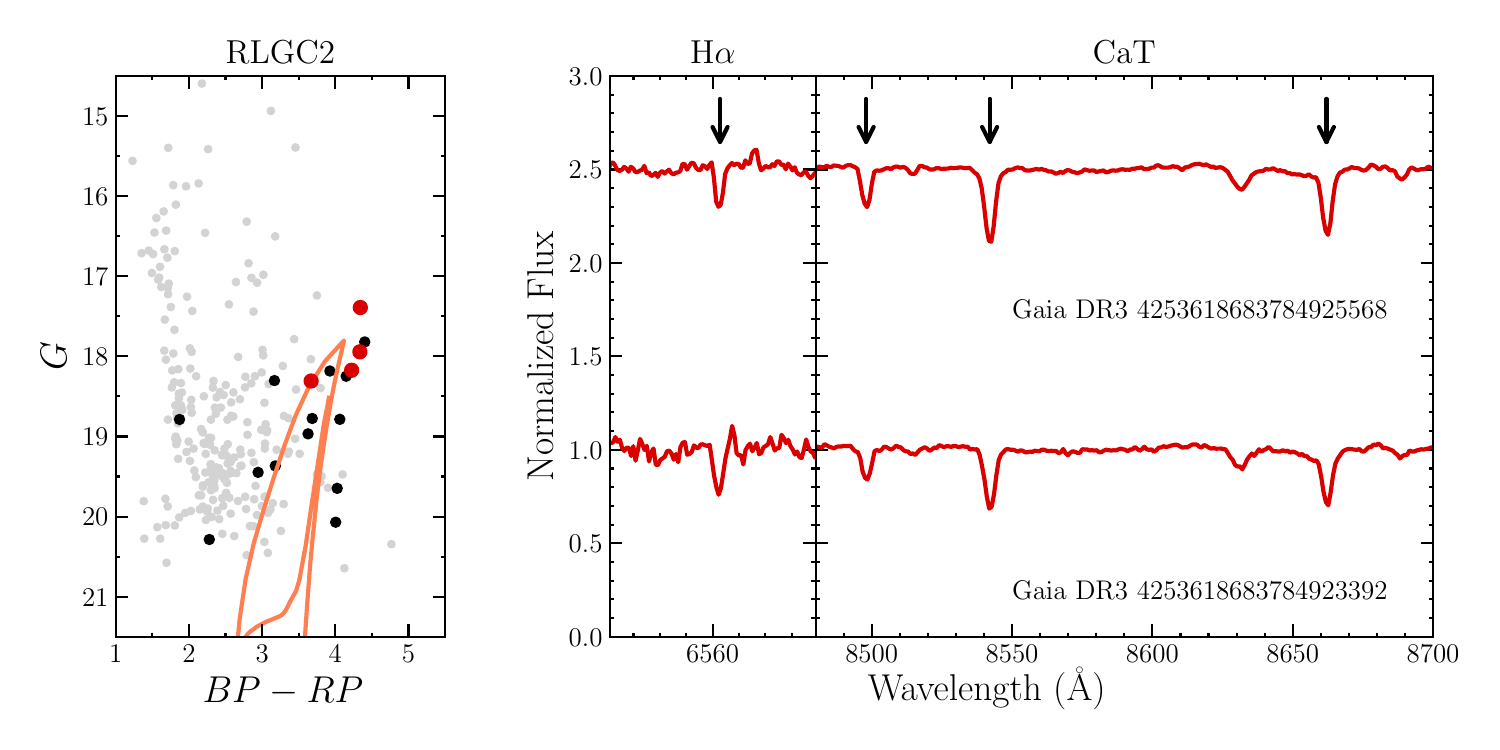}
\caption{Same as Fig. \ref{fig:data_Pfleiderer2}, but Mu\~noz 1 (green) and RLGC2 (red). Reference BaSTI isochrone are 12 Gyr, $\feh = -1.4$ dex and $\feh = -2.3$ dex, respectively, and $\alpha$-enanched ($\afeh = +0.4$ dex), assuming distance and reddening from \citep{munoz2012,ryu2018}. Stars in our sample identified as outliers in $\vlos$ are marked with a coloured cross, if present.}
\label{fig:data_3}%
\end{figure*}
%-------------------------------------------------------------
In Table \ref{tab:obs_stars}, we report the observational information for targeted stars. In Figs. \ref{fig:data_2} - \ref{fig:data_3}, we present the \textit{Gaia} CMDs of proper motion selected members of Koposov 1, Koposov 2, \ref{fig:data_2} Mu\~noz 1, and RLGC2 \citep[$p>80\%$,][]{vasiliev&baumgardt2021}. Also, we plot a portion of two stellar spectra per cluster, close to the H$\alpha$ and CaT lines.
%-------------------------------------------------------------
\section{Output of the BGM model}\label{app:besa}
%-------------------------------------------------------------
\begin{figure*}
\centering
\begin{minipage}{0.4\textwidth}
\centering        
\includegraphics[width=0.8\textwidth]{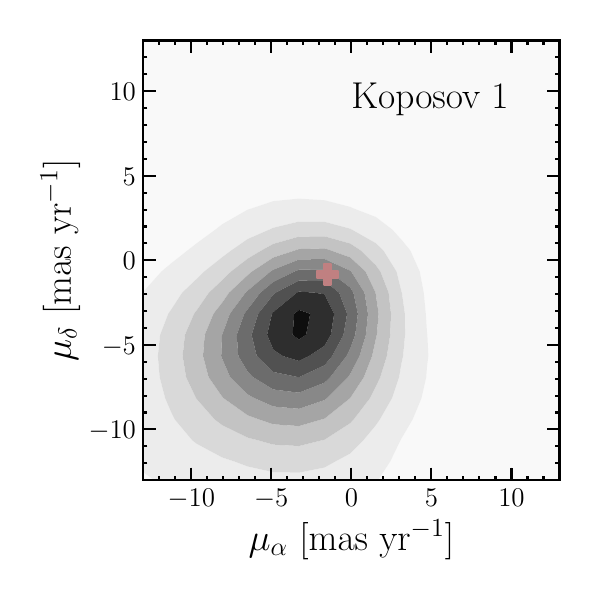}  
\includegraphics[width=1.0\textwidth]{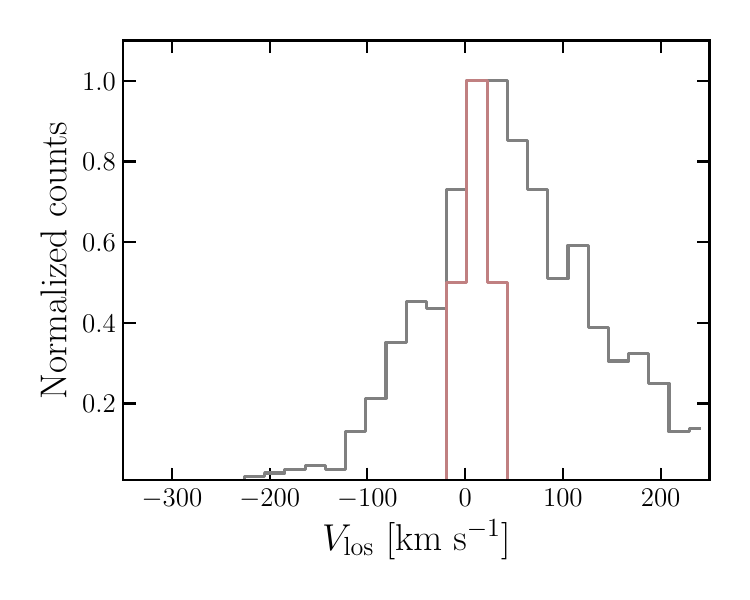}           
\includegraphics[width=0.8\textwidth]{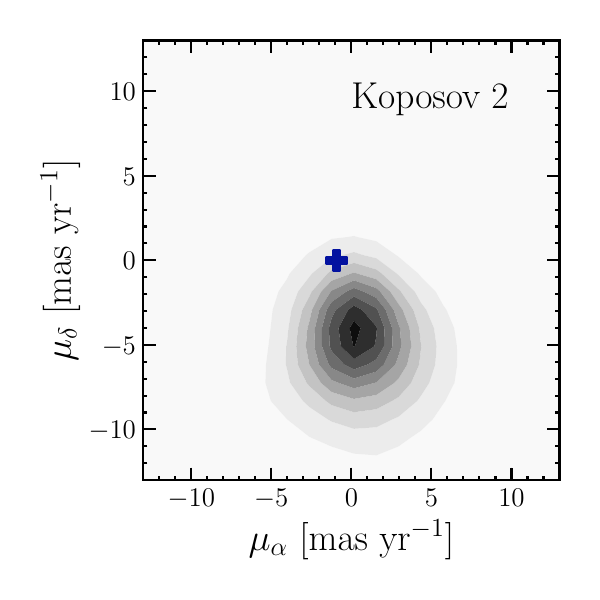}  
\includegraphics[width=1.0\textwidth]{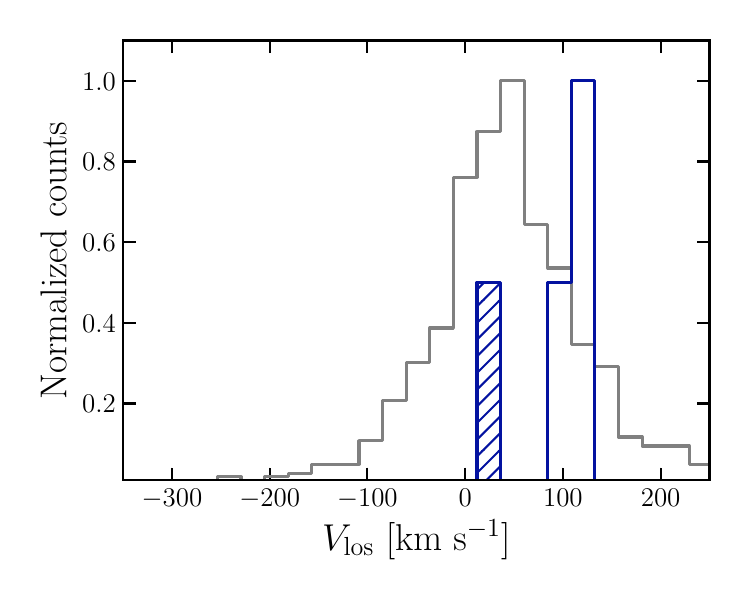}  
\end{minipage}
\begin{minipage}{0.4\textwidth}
\centering
\includegraphics[width=1.0\textwidth]{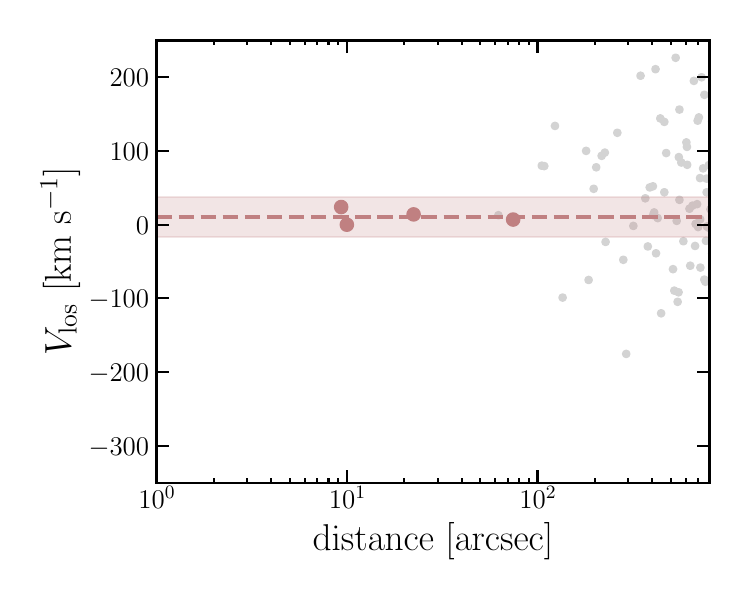} 
\includegraphics[width=1.0\textwidth]{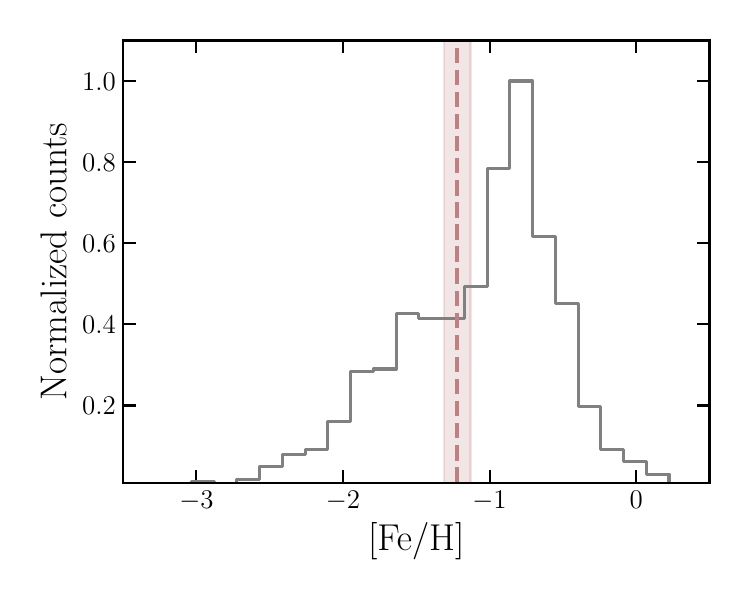} 
\includegraphics[width=1.0\textwidth]{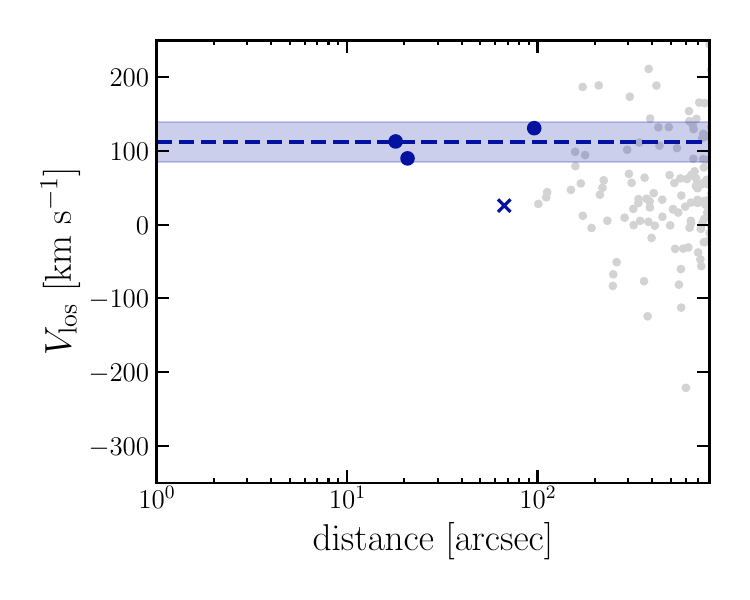} 
\includegraphics[width=1.0\textwidth]{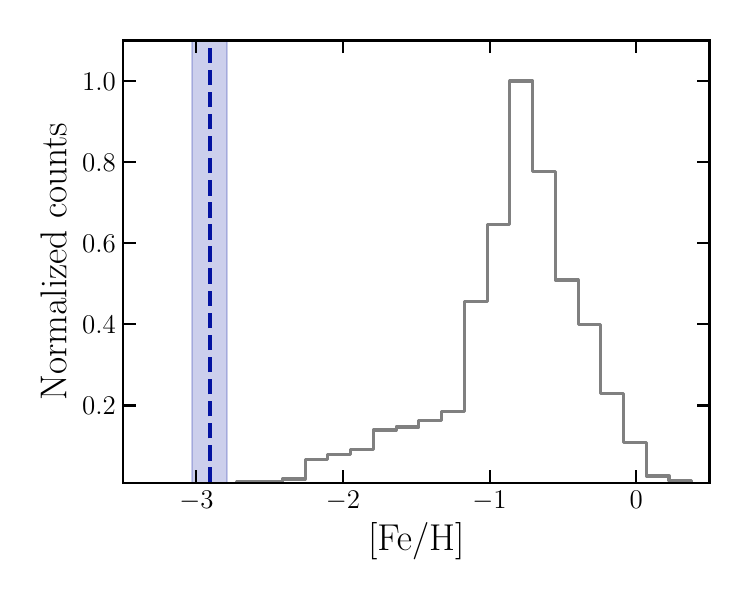}  
\end{minipage}  
\caption{Same as Fig. \ref{fig:besa_pwm2} for Koposov 1 and Koposov 2. The histogram bin containing the velocity outlier is highlighted with diagonal hatching. For the clusters for which the metallicity is measured for a single star, we plot the corresponding value, and the shaded area represents the associated 1$\sigma$ uncertainty.}
\label{fig:app_besa1}%
\end{figure*}
%-------------------------------------------------------------
\begin{figure*}
\centering
\begin{minipage}{0.4\textwidth}
\centering        
\includegraphics[width=0.8\textwidth]{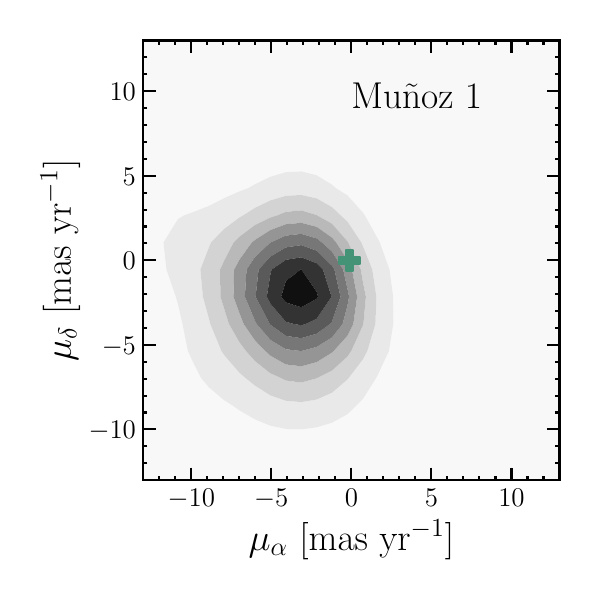}  
\includegraphics[width=1.0\textwidth]{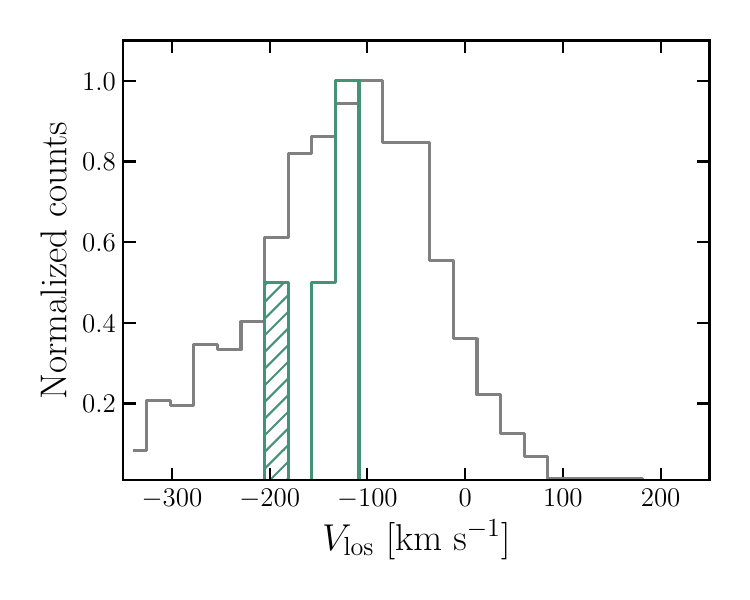}           
\includegraphics[width=0.8\textwidth]{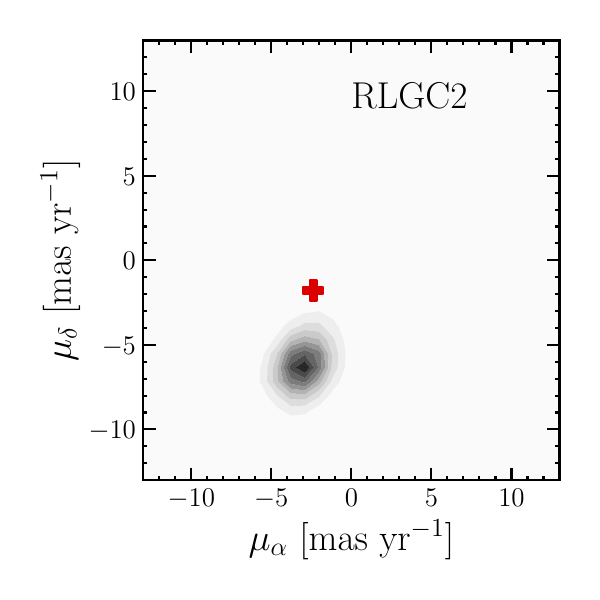}  
\includegraphics[width=1.0\textwidth]{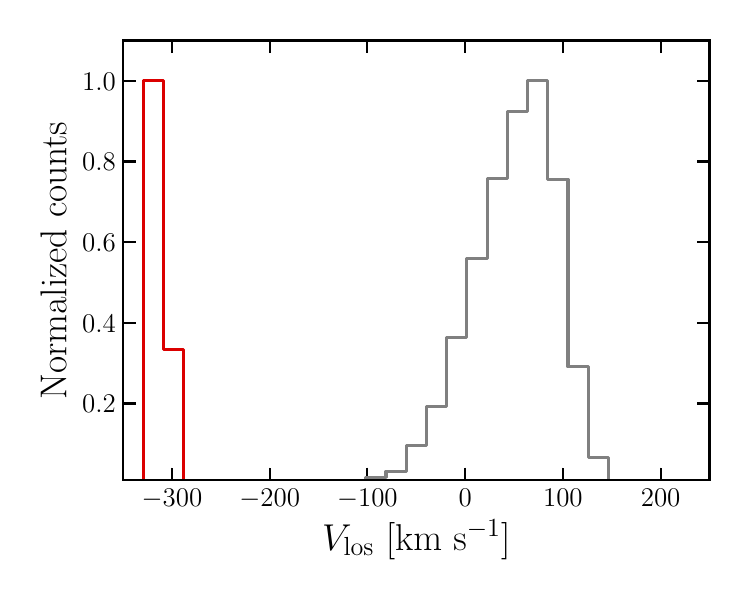}  
\end{minipage}
\begin{minipage}{0.4\textwidth}
\centering
\includegraphics[width=1.0\textwidth]{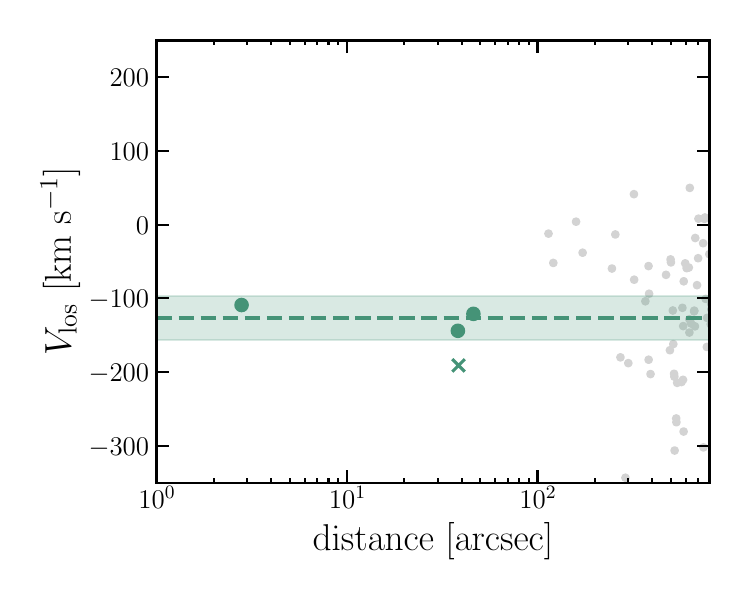} 
\includegraphics[width=1.0\textwidth]{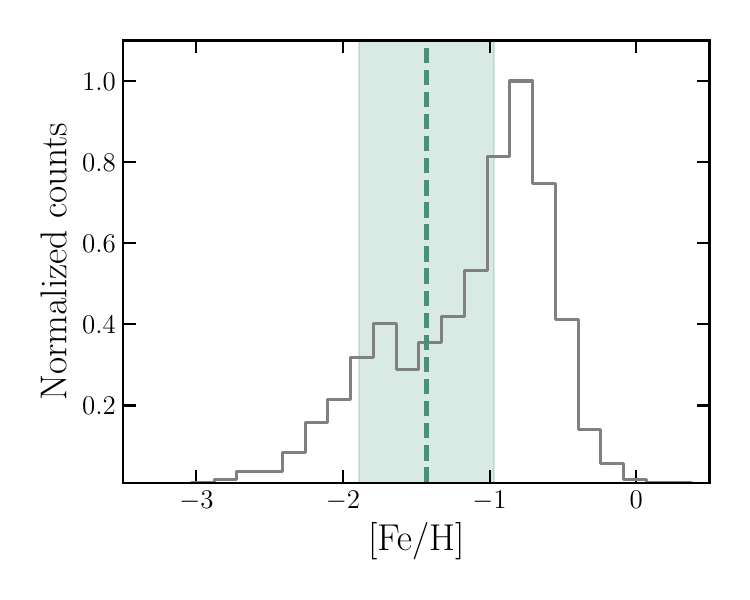} 
\includegraphics[width=1.0\textwidth]{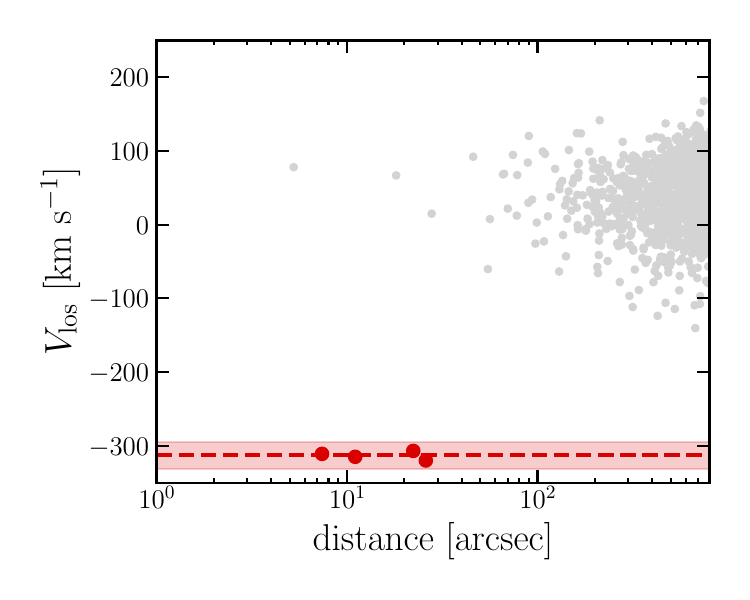} 
\includegraphics[width=1.0\textwidth]{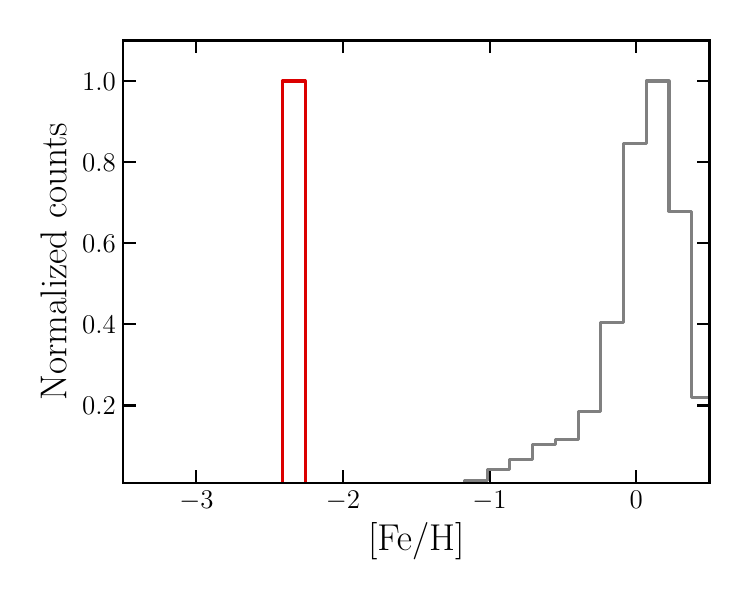}  
\end{minipage}  
\caption{Same as Fig. \ref{fig:app_besa1} for Mu\~noz 1 and RLGC2.}
\label{fig:app_besa2}%
\end{figure*}

In Fig. \ref{fig:app_besa1} - \ref{fig:app_besa2}, we present the additional output of the BGM model for Koposov 1, Koposov 2, Mu\~noz 1, and RLGC2. We note that for the first three clusters the spatial window around each cluster center was expanded to $2^\circ$ to enhance the statistics of the simulated field populations. As the metallicity is available for only one star in Koposov 1, Koposov 2, and Mu\~noz 1, the BGM metallicity distribution is compared with the measured value and its $1\sigma$ uncertainty.

%-------------------------------------------------------------
\section{The impact of different distance estimates on the chemodynamical analysis}\label{app:dist}
%-------------------------------------------------------------
In Section \ref{sec:analysis}, we adopted a set of homogeneous distance estimates by \citet{paust2014} for Koposov 1 and Koposov 2, namely $34.9\pm1.6$ kpc and $33.3\pm1.5$ kpc, respectively. However, the literature reports significantly discrepant values for these two systems: in the discovery paper by \citet{koposov2007}, distances of $\sim50$ kpc for Koposov 1 and $\sim40$ kpc for Koposov 2 were estimated, while more recent works suggest 48.3 kpc for Koposov 1 \citep{munoz2018} and 24 kpc for Koposov 2 \citep{geha2026}.

Here we briefly assess how adopting these alternative distances affects both the dynamical properties of the clusters and the inferred metallicities. If Koposov 1 is placed at 48.3 kpc, it would become dynamically compatible with an association with the Cetus progenitor, as proposed by \citet[][eDR3 edition]{massari19}, and the derived metallicity becomes $\sim0.2$ dex more metal-poor. Koposov 2, on the other hand, remains high energy and unassociated for all the distances considered, although larger assumed distances naturally move it at higher orbital energies, $L_{\mathrm{z}}$, and $L_{\perp}$. We note that adopting a distance of 24 kpc would make its RGB star $\sim0.15$ dex more metal-rich.
%-------------------------------------------------------------
\section{Comparison with different CaT metallicity relations}\label{app:met}
%-------------------------------------------------------------
To verify and assess the robustness of our results, as well as their sensitivity to the choice of calibration relations, we also derived stellar metallicities using three additional calibration relations. In Table \ref{tab:app_met}, we report the different metallicity values obtained applying the relations from \citet{carrera2007}, \citet{starkenburg2010} and \citet{carrera2013}. Given that such relations depends on $M_{V}$, we computed magnitudes the $V$ band starting from \textit{Gaia} DR3 photometry \citep{GaiaDR3} and using the empiric transformations from \citet{riello2021}. To place all estimates on an homogeneous scale, we also apply the \citet{navabi2026} calibrations using the $M_{V}$ absolute magnitude.

We do not detect any significant systematic offset between the metallicity obtained with the $M_{G}$ and $M_{V}$ calibrations presented by \citet{navabi2026}, with the only notable exception being RLGC2, for which the metallicity derived using the $V$-band magnitude is approximately 0.2 dex higher.

More generally, the \citet{navabi2026} relations tend to yield, on average, slightly more metal-poor metallicities for all stars in our sample, with offsets of $\sim0.1-0.3$ dex relative to the other three adopted calibrations. This is particularly evident for Pfleiderer 2, which is systematically more metal-poor when adopting the \citet{navabi2026} calibration. The differences reach up to $\sim0.3$ dex, being inconsistent at $>2.2\sigma$ compared to the other three estimates, which instead provide compatible values. Such inconsistencies are expected in the high metallicity regime (down to $\feh \sim -1.5$ dex) and likely reflect the different zero-points of the calibrations. In particular, these relations are anchored at the metal-rich end using the open cluster NGC 6791; however \citet{navabi2026} adopt a more recent spectroscopic determination of its metallicity, which is lower than the value assumed in earlier works \citep{carrera2007,carrera2013}. This naturally shifts the calibration toward slightly more metal-poor values, especially at the metallicity of Pfleiderer 2 .

In the end, we find that the \citet{carrera2007} calibration provides a metallicity that is $\sim 0.6$ dex more metal-rich for Koposov 2, while the other three relations provide metallicities that are consistent within the uncertainties. This is possibly due to the fact that Koposov 2 lies in a metallicity regime more metal-poor than the range over which the \citet{carrera2007} relation is well constrained (i.e. down to $\feh \sim -2.2$ dex). 

%-------------------------- Table -------------------------
\begin{table*}
\caption{Metallicity obtained using different scaling relations and $M_{V}$ as a proxy for the luminosity.}        
\label{tab:app_met}      
\centering          
\begin{tabular}{lccccc}  
\hline 
\hline      
Cluster / \textit{Gaia} DR3 & $\mathrm{[Fe/H]_{\mathrm{C07}}}$ & $\mathrm{[Fe/H]_{\mathrm{S10}}}$ & $\mathrm{[Fe/H]_{\mathrm{C13}}}$ & $\mathrm{[Fe/H]_{\mathrm{N26,V}}}$ & Member \\ 
 & (dex) & (dex) & (dex) & (dex) & \\ 
\hline 
3919839462184529664 & $-$1.01$\pm$0.07 & $-$1.03$\pm$0.21 & $-$1.09$\pm$0.09 & $-$1.24$\pm$0.10 & y \\
3919839767126601728 & - & - & - & - &  y \\
3919839767126647680 & - & - & - & - &  y \\
3919839771422278016 & - & - & - & - &  y \\
Koposov 1 &  &  &  &  & \\
\hline 
874275674394169984 & $-$2.34$\pm$0.06 & $-$2.82$\pm$0.17 & $-$2.84$\pm$0.11 & $-$2.91$\pm$0.13  & y  \\
874275609971796992 & - & - & - & - & y  \\
874269730159393536 & - & - & - & - & y  \\
874275850488194176 & - & - & - & - & n  \\
Koposov 2          &  &  &  &  & \\
\hline 
1693370742839913088 & - & - & - & - & y \\
1693394146117710592 & - & - & - & - & y \\
1693394176181484672 & $-$1.26$\pm$0.38 & $-$0.98$\pm$0.49 & $-$1.23$\pm$0.47 & $-$1.42$\pm$0.43  & y \\
1693393420267235840 & - & - & - & - & n \\
Mu\~noz 1           &  &  &  &  &  \\
\hline 
4172573170590468992 & $-$0.48$\pm$0.23 & $-$0.30$\pm$0.33 & $-$0.40$\pm$0.22 & $-$0.67$\pm$0.21 & y \\
4175575730690574208 & $-$0.59$\pm$0.19 & $-$0.51$\pm$0.31 & $-$0.53$\pm$0.18 & $-$0.78$\pm$0.18 & y \\
4172549561149372160 & $-$0.69$\pm$0.23 & $-$0.43$\pm$0.32 & $-$0.64$\pm$0.22 & $-$0.86$\pm$0.20 & y \\
4172549836027324800 & $-$0.55$\pm$0.18 & $-$0.48$\pm$0.30 & $-$0.52$\pm$0.17 & $-$0.77$\pm$0.16 & y \\
4172549840328133120 & $-$0.35$\pm$0.19 & $-$0.27$\pm$0.32 & $-$0.29$\pm$0.18 & $-$0.58$\pm$0.18 & y \\
4175575863825402880 & $-$0.43$\pm$0.17 & $-$0.39$\pm$0.29 & $-$0.41$\pm$0.16 & $-$0.68$\pm$0.15 & y \\
4172573269368904448 & $-$0.55$\pm$0.19 & $-$0.56$\pm$0.30 & $-$0.52$\pm$0.19 & $-$0.77$\pm$0.17 & y \\
Pfleiderer 2 &        $-$0.51$\pm$0.07 & $-$0.42$\pm$0.12 & $-$0.47$\pm$0.07 & $-$0.73$\pm$0.07 & \\
\hline 
4253618683784925568 & $-$1.91$\pm$0.23 & $-$1.91$\pm$0.23 & $-$2.10$\pm$0.18 & $-$2.05$\pm$0.13 & y \\
4253618683784927872 & $-$1.89$\pm$0.19 & $-$1.92$\pm$0.22 & $-$2.12$\pm$0.17 & $-$2.08$\pm$0.13 & y \\
4253618683784923392 & $-$2.03$\pm$0.22 & $-$2.02$\pm$0.22 & $-$2.25$\pm$0.18 & $-$2.18$\pm$0.13 & y \\
4253618683784929536 & $-$1.97$\pm$0.19 & $-$1.95$\pm$0.22 & $-$2.22$\pm$0.17 & $-$2.17$\pm$0.13 & y \\
RLGC2               & $-$1.95$\pm$0.10 & $-$1.95$\pm$0.11 & $-$2.17$\pm$0.09 & $-$2.12$\pm$0.07 & \\
\hline 
\end{tabular}
\end{table*}
%-------------------------------------------------------------

\end{appendix}

\end{document}